\documentclass[a4paper, superscriptaddress, notitlepage, amsmath,amssymb, aps, twocolumn, prl, floatfix, 10pt]{revtex4-2}
\usepackage{graphicx}
\usepackage{bm,bbm}
\usepackage[urlcolor=blue]{hyperref}
\hypersetup{colorlinks=true,allcolors=blue}
\usepackage{amssymb} 
\usepackage{todonotes}

\newcommand{\rmd}{\mathrm{d}}

\newcommand{\btheta}{\boldsymbol{\theta}}

\usepackage{color}\usepackage{xcolor}
\colorlet{mylinkcolor}{blue!66!black!80}
\definecolor{grey}{rgb}{0.6,0.6,.6}
\definecolor{darkgrey}{rgb}{0.4,0.4,.4}
\definecolor{darkgreen}{rgb}{0,0.4,0}
\definecolor{lightgreen}{rgb}{0,0.7,0}
\definecolor{darkred}{rgb}{0.5,0,0}
\newcommand{\red}[1]{{\color{red}#1}}

\newcommand{\blue}[1]{{\color{blue}#1}}

\newcommand{\smfig}[2]{\IfFileExists{#1}{\includegraphics[width=#2]{#1}}%
  {\fbox{\parbox[c][4cm][c]{#2}{\centering\red{Figure to be produced: \texttt{\detokenize{#1}}}}}}}

\renewcommand{\blue}[1]{#1}

\begin{document}

\title{Uncovering nonlinear dynamics \blue{hidden by correlated noise}}

\author{Cai Dieball}
\affiliation{Institute for Theoretical Physics, University of Cologne,  50937 K\"oln, Germany}

\author{Leonardo Rydin Gorj\~ao}
\affiliation{Department of Environmental Sciences, Open University of the Netherlands, Heerlen, The Netherlands}
\affiliation{Institute of Physics, Norwegian University of Life Sciences, \AA s, Norway}

\author{Klaus Lehnertz}
\affiliation{Department of Epileptology, University of Bonn Medical Center, Venusberg Campus 1, Bonn, Germany}
\affiliation{Helmholtz Institute for Radiation and Nuclear Physics, University of Bonn, Nussallee 14-16, Bonn, Germany}
\affiliation{Interdisciplinary Centre for Complex Systems, University of Bonn, Brühler Str. 7, Bonn, Germany}

\author{Dirk Witthaut}
\affiliation{Institute for Theoretical Physics, University of Cologne,  50937 K\"oln, Germany}
\affiliation{Institute of Climate and Energy Systems: Energy Systems Engineering (ICE-1), Forschungszentrum J\"ulich, 52425 J\"ulich, Germany}

\begin{abstract}
The drift of a stochastic system, i.e., the deterministic forces acting on it, is commonly inferred from trajectories via the mean increment at a given state, known as the Kramers--Moyal coefficient. This approach implicitly assumes that the noise is uncorrelated in time, which is rarely the case in real-world complex systems. We show that temporally correlated noise can hide the drift entirely from short increments, since for stationary one-dimensional dynamics with differentiable trajectories the mean velocity at any given position vanishes. Finite sampling intervals partially restore the drift, which explains the partial success of past applications. To gain intuition, we trace this effect to Onsager's regression of fluctuations. Finally, we introduce a two-increment estimator that eliminates the noise correlations and infers the drift together with the noise correlation time. We demonstrate it on a simulated double well, on a tracer in an active bacterial bath, and on frequency fluctuations in the Great Britain power grid, where it reveals important nonlinear signatures of frequency control despite long and intricate correlations in power imbalances.
\end{abstract}
\maketitle

Inferring forces from motion is a fundamental inverse problem in physics. In complex systems, ranging from living matter to the climate, the effective equations of motion can rarely be derived from first principles. Instead, they must be inferred from fluctuating trajectories by reconstructing a stochastic model whose deterministic part, the drift $f(x)$, encodes the restoring forces and control mechanisms of interest~\cite{friedrich2011approaching,Peinke2019,RahimiTabar2019}.
A particularly simple and widely used approach is to estimate the mean displacement over a short time interval $\Delta t$, conditioned on the present state $x$,
\begin{equation}
  \mathrm{KM}(x,\Delta t)\equiv
  \frac{\langle x_{t+\Delta t}-x_t\,|\,x_t=x\rangle}{\Delta t}
  \ \xrightarrow{\ \Delta t\to 0\ }\ f(x)\,,
  \label{eq:KM}
\end{equation}
whose limit yields the first Kramers--Moyal (KM) coefficient for Markov diffusion processes~\cite{siegert1998analysis,gardiner2004handbook,oksendal2003stochastic}.
The conditional average can be estimated directly from stationary trajectories, making the method intuitive and computationally efficient. Accordingly, KM estimation has found applications across disciplines, from turbulence~\cite{Friedrich1997,Friedrich2020} and financial markets~\cite{Friedrich2000} to biological motion~\cite{Brueckner2019}, physiological signals~\cite{Prusseit2007,Lamouroux2009,Hashtroud2019}, paleoclimate records~\cite{Ditlevsen1999,Riechers2025}, and renewable energy systems~\cite{Milan2013,rydin2020data,RydinGorjao2020}. Closely related approaches are widely used in econometrics~\cite{Stanton1997,AtSahalia1996,Bandi2003} and stochastic force inference in biophysics~\cite{Frishman2020,ferretti2020building,Brueckner2020}.

\blue{All these applications rest on an assumption that is rarely stated: the noise must be uncorrelated in time, known as \textit{white}, so that it averages out when conditioned on the present state.
In real-world complex systems, however, noise is never truly white.
It typically arises from unobserved degrees of freedom with finite relaxation times, as formalized by Mori and Zwanzig~\cite{Zwanzig1961,Mori1965}, and its correlation time is often not short compared with the time scales on which the system is observed.
For example, swimming bacteria exert temporally correlated stochastic forces on suspended tracers~\cite{Wu2000,Maggi2014,VillalobosConcha2025}, viscoelastic baths exhibit memory effects~\cite{Ginot2022}, slow climate variables are driven by weather fluctuations that persist over days~\cite{Hasselmann1976,frankignoul1985sea,Morr2024,Morr2025}, and power-grid frequency is driven by imbalances between electricity supply and demand that are correlated over minutes~\cite{milano2013systematic,roberts2016validation,tyloo2021primary,gorbunov2022inertia}.
While much effort has been devoted to correcting KM estimates for finite sampling intervals~\cite{Ragwitz2001,Friedrich2002,Sura2002,Gottschall2008,Anteneodo2009,Lade2009,Anteneodo2010,Honisch2011,RydinGorjao2021} and measurement noise~\cite{Boettcher2006,Lehle2011}, temporal correlations in the noise pose a fundamentally different challenge.}

\begin{figure}
\centering
\includegraphics[width=\columnwidth]{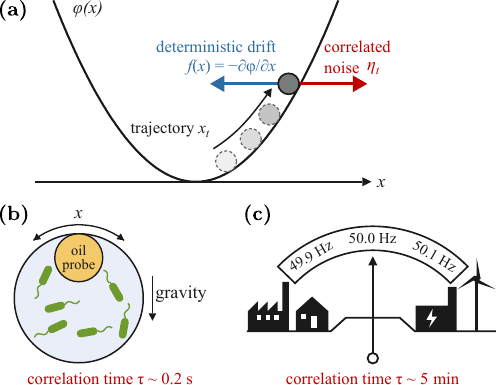}
\caption{
(a) The goal is to reconstruct the drift $f(x)$ in the presence of a correlated noise $\eta_t$. Consider a coordinate at position $x$ at the edge of a confining potential $\varphi(x)$. While the drift pulls the system towards the potential minimum, the correlated noise that displaced it is still acting. The conditional mean $\langle\eta_t|x_t=x\rangle\ne0$ obscures the inference of $f(x)$.
(b) Biological application: An oil droplet moves in an active bacterial bath~\cite{VillalobosConcha2025}. Bacterial motion generates correlated noise with time scale $\tau\approx 0.2$\,s, while gravity provides a restoring force.
(c) Engineering application: Power-grid frequency is driven by imbalances between electricity generation and demand~\cite{RydinGorjao2020}. These imbalances introduce stochastic noise with correlation times on the order of minutes, while load-frequency control provides the restoring drift.
}
\label{fig:fig1}
\end{figure}

Why does correlated noise complicate the inference of the deterministic part of the dynamics?
Consider a coordinate $x_t$ subject to a restoring drift $f(x)$ and stochastic noise $\eta_t$, as illustrated in Fig.~\ref{fig:fig1}.
Suppose the coordinate is found on the slope of a confining potential, away from its minimum.
For memoryless noise, the random fluctuations that displaced the system do not persist into the next increment, and the mean motion reveals the restoring drift, as expressed by Eq.~\eqref{eq:KM}.
For correlated noise, however, the noise that brought the system away from the minimum is still acting due to correlations in time~\cite{Lehle2018,Willers2021}.
Consequently, the noise does not average to zero at a given position, $\langle\eta_t|x_t=x\rangle\neq 0$, and its contribution obscures the deterministic drift. 
How severely does correlated noise compromise drift inference, both in theory and in practical applications? Can this failure be understood from the physics of fluctuations and their relaxation? And can we develop a robust \blue{extension of} the Kramers--Moyal estimation that recovers the underlying drift in the presence of correlated noise?

\blue{In this Letter, we address these questions by showing that correlated noise causes the KM estimate to vanish completely as $\Delta t\to0$, a failure we also explain with Onsager's regression hypothesis.
We then introduce a two-increment estimator that recovers the drift and the noise correlation time directly from data, and demonstrate it on simulations, experimental data from an active bacterial bath, and power-grid frequency recordings.}

\paragraph*{Correlation masks the drift.}
We consider a one-dimensional system subject to random noise $\eta_t$,
\begin{equation}
    \dot x_t=f(x_t)+\eta_t\,,
    \label{eq:model}
\end{equation}
where $f(x)$ denotes the deterministic drift.
We assume that $\eta_t$ \blue{is correlated in time and has continuous sample paths}, and that the system reaches a stationary state with vanishing probability current.

These assumptions already have a remarkable consequence for stochastic inference.
Since $\eta_t$ is continuous, $x_t$ is differentiable.
For any smooth function $g(x)$ with antiderivative $G(x)$, stationarity implies that $\langle G(x_t)\rangle$ is constant over time, and consequently
$0=\frac{\mathrm{d}}{\mathrm{d}t}\langle G(x_t)\rangle
    =\langle g(x_t)\dot x_t\rangle$. 
Since $g$ is arbitrary, we obtain (see also Ref.~\cite{Lehle2018})
\begin{equation}
    \langle\dot x_t\,|\,x_t=x\rangle=0
    \quad\Rightarrow\quad
    \langle\eta_t\,|\,x_t=x\rangle=-f(x)\,.
    \label{eq:zero_current}
\end{equation}
Thus, the noise correlation exactly compensates the deterministic drift on average at every position. 
Consequently, the Kramers--Moyal estimator in Eq.~\eqref{eq:KM} vanishes in the limit $\Delta t\to0$, irrespective of the drift $f(x)$. 
\blue{That is, we obtain $\mathrm{KM}(x,\Delta t)\to0$ for $\Delta t\to 0$, drastically contradicting the common use of Eq.~\eqref{eq:KM}.}

Remarkably, this result does not depend on the statistics or the correlation function of the random noise, but follows from stationarity and differentiable trajectories alone.
White noise constitutes a singular exception, since its trajectories are not differentiable.

\paragraph*{Finite sampling partially restores the drift.} 
To quantify the effect at finite sampling intervals, 
\blue{we have to specify a model for the correlated noise.} 
We consider the important case of Ornstein--Uhlenbeck (OU) noise~\cite{Uhlenbeck1930,Lehle2018},
\begin{equation}
    \mathrm{d}\eta_t=-\frac{\eta_t}{\tau}\,\mathrm{d}t
    +\frac{\sqrt{2D}}{\tau}\,\mathrm{d}W_t\,,
    \label{eq:OU}
\end{equation}
where $\tau>0$ denotes the correlation time, $D>0$ characterizes the noise strength, and $W_t$ is a Wiener process.
The stationary noise autocorrelation decays exponentially,
$\langle\eta_t\eta_{t'}\rangle=(D/\tau)e^{-|t-t'|/\tau}$.
\blue{OU noise is a direct extension of white noise: it adds a single parameter, the memory time $\tau$, and recovers the standard white-noise picture behind KM estimation as $\tau\to0$, while capturing for $\tau>0$ the effects of correlations that white noise neglects by construction.
Consequently}, OU noise is widely used to model correlated fluctuations in active matter, viscoelastic systems, climate dynamics, and power grids~\cite{Fodor2016,Martin2021,Ginot2022,Morr2024,Morr2025,milano2013systematic,tyloo2021primary}.

\blue{In light of the concrete model in Eq.~\eqref{eq:OU}, we can now compare the dependence on sampling interval $\Delta t$ with the correlation time scale $\tau$.} 
For finite increments, the noise gradually decorrelates, such that
$\langle\eta_{t+s}|x_t=x\rangle
=e^{-s/\tau}\langle\eta_t|x_t=x\rangle
=-e^{-s/\tau}f(x)$ for any $s>0$. 
\blue{Expanding the increment to second order in }
$\Delta t$ (see \footnote{See Supplemental Material (yet to be published) for derivations, implementation details, and details on the data, including references \cite{Carlstein1986,Gillespie1996,Willers2021zenodo}.}), we obtain
\begin{equation}
    \mathrm{KM}(x,\Delta t)
    =\frac{\Delta t}{2\tau}\,f(x)
    +\mathcal{O}(\Delta t^2)\,.
    \label{eq:km_first}
\end{equation}
Thus, finite sampling intervals partially restore the inferred drift, but its amplitude is suppressed by a factor $\Delta t/(2\tau)$ to leading order. \blue{This explains partly successful applications of the estimator in Eq.~\eqref{eq:KM}, where qualitative inference of $f(x)$ was achieved despite having $\Delta t<\tau$ \cite{RydinGorjao2025,wen2025nonstandard}.}

\begin{figure*}
  \centering
  \includegraphics[width=.98\textwidth]{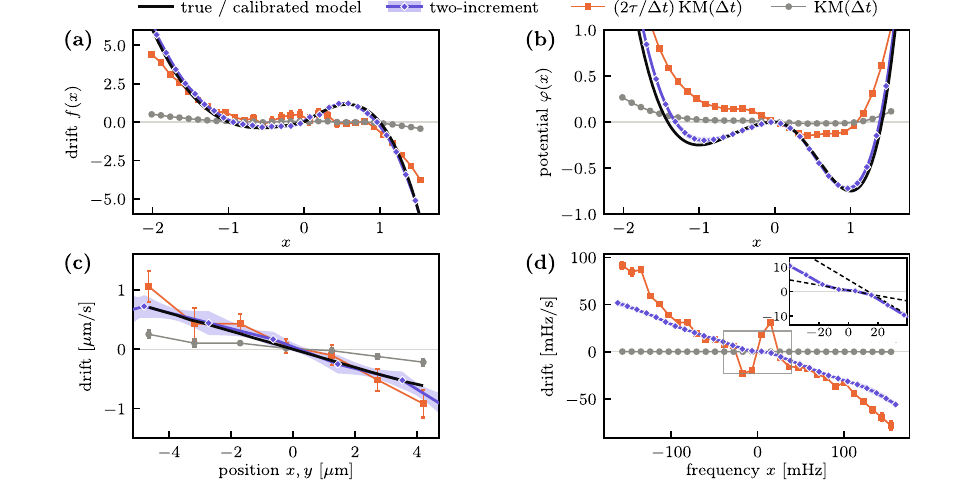}
    \caption{
    \blue{Drift recovered by the two-increment estimator [purple, Eq.~\eqref{eq:exact}], compared with
$\mathrm{KM}(\Delta t)$ [gray, Eq.~\eqref{eq:KM}] and its first-order correction
$(2\tau/\Delta t)\,\mathrm{KM}(\Delta t)$ [orange, Eq.~\eqref{eq:km_first}]. 
(a)~Drift and (b)~potential $\varphi(x)=-\int_0^x f(x')\rmd x'$ of a simulated asymmetric double well, $f(x)=x-x^3$ for $x<0$ and $f(x)=3(x-x^3)$ for $x>0$ (black; $\tau=0.4$, $D=4$, $\Delta t=0.1$, $5\times10^5$ samples; inferred $\hat\tau=0.44$). (c)~Oil probe in an active bacterial bath with trajectories from Ref.~\cite{VillalobosConcha2025} ($\Delta t=0.1$\,s; $\hat\tau=0.21$\,s), with the drift of the calibrated model of Ref.~\cite{VillalobosConcha2025}. 
(d)~Great Britain (GB) power-grid frequency \cite{Note2}, one year at $\Delta t=1$\,s ($\hat\tau\approx5$\,min).
Inset: zoom into $|x|\le40$\,mHz, with linear fits to middle ($|x|\le15$\,mHz) and right ($x\ge20$\,mHz) parts of the drift (black, dashed) revealing a pronounced nonlinearity.
See \cite{Note1} for inference of each $\hat\tau$, description and discussion of error bars, and details on the experimental data and calibration.}}
  \label{fig:fig2}
\end{figure*}

This result has important consequences for stochastic inference.
Decreasing the sampling interval does not improve the KM estimate but instead drives it towards zero.
Indeed, vanishing KM coefficients at short time scales have previously been associated with a finite Markov--Einstein time~\cite{Honisch2011,Honisch2012}, and the established remedy is to discard short increments~\cite{Lueck2006,Stresing2011,friedrich2011approaching}.
However, this approach is fragile: 
The sampling interval must be sufficiently long for the noise to decorrelate, but still short enough to resolve the deterministic dynamics.
Such a separation of time scales is often absent.
Furthermore, Eq.~\eqref{eq:km_first} shows that finite-time estimates may reproduce the shape of the drift while severely underestimating its magnitude.
Even correcting for the leading-order suppression factor $ \Delta t/(2\tau) $ does not generally resolve the problem, as higher-order corrections can substantially distort the inferred drift. 
\blue{Fig.~\ref{fig:fig2} shows that both the KM estimate and its leading-order correction can fail to recover the drift.}

\paragraph*{Inference and Onsager's regression.} 
Further physical insight is provided by Onsager's regression of fluctuations~\cite{onsager1931reciprocal1,onsager1931reciprocal2,reimann2024onsager}, which relates the relaxation of spontaneous fluctuations to macroscopic dynamics.
For linear drift $f(x)=-ax$ and stationary Gaussian noise, $x_t$ is a stationary Gaussian process.
Assuming $\langle x\rangle=0$ and defining the autocorrelation function $C(\Delta t)=\langle x_{t+\Delta t}x_t\rangle$, the conditional mean increment satisfies the exact relation (see \cite{Note1} and Ref.~\cite{reimann2024onsager})
\begin{equation}
  \langle x_{t+\Delta t}-x_t\,|\,x_t=x\rangle
  =-\left[1-\frac{C(\Delta t)}{C(0)}\right]x\,.
  \label{eq:onsager}
\end{equation}
This relation holds irrespective of the correlation time of the noise.
However, it does not imply that spontaneous fluctuations relax according to the deterministic drift $\dot x=-ax$ alone.
As shown above, a fluctuation to position $x$ is accompanied by a correlated noise with conditional mean $\langle\eta_t|x_t=x\rangle=ax$, which initially compensates the restoring drift.
The subsequent regression is therefore influenced by the relaxation of the hidden noise.
In Onsager's picture, the macroscopic description must retain the relevant slow variables, while rapidly relaxing degrees of freedom can be eliminated.
Correlated noise introduces precisely such a hidden slow variable, which cannot be neglected when interpreting the regression of fluctuations.

The distinction becomes particularly clear in the limit of short time increments.
For white noise, the autocorrelation function $C(\Delta t)=C(0)e^{-a|\Delta t|}$ has a cusp at $\Delta t=0$, such that $C(\Delta t)/C(0)=1-a\Delta t+\mathcal{O}(\Delta t^2)$ for $\Delta t>0$.
Equation~\eqref{eq:onsager} then recovers the conventional KM result, $\mathrm{KM}(x,\Delta t)\to-ax=f(x)$.
For correlated noise with differentiable trajectories, however, stationarity implies $\dot C(0)=\langle\dot x_t x_t\rangle=0$, and hence $C(\Delta t)/C(0)=1+\mathcal{O}(\Delta t^2)$.
Consequently, the KM estimator vanishes as $\Delta t\to0$, in agreement with our general result.

\paragraph*{A robust two-increment estimator.}
We now develop an estimator that recovers the drift $f(x)$ despite correlated noise.
\blue{
Building on previous approximate approaches~\cite{Lehle2018,Willers2021}, the key idea is to exploit the temporal correlations of the noise rather than attempting to avoid them. For the OU model in Eq.~\eqref{eq:OU}, the correlated contribution of the hidden noise can be eliminated by combining two consecutive increments. 
The precise relation to Refs.~\cite{Lehle2018,Willers2021} and to approaches for underdamped dynamics~\cite{Lehle2015,Brueckner2019,Brueckner2020} and partially observed diffusions~\cite{Dembo1986,Campillo1989,Golightly2008} is explained in~\cite{Note1}.} 
Integrating Eq.~\eqref{eq:model} over sampling intervals $[t_n,t_{n+1}]$, with $t_n=n\Delta t$ \blue{and $x_n\equiv x_{t_n}$}, yields the exact relation (see \cite{Note1})
\begin{equation}
  y_n\equiv\Delta x_n-\rho\,\Delta x_{n-1}
  =F_n-\rho F_{n-1}+\zeta_n\,,
  \label{eq:exact}
\end{equation}
where $\rho=e^{-\Delta t/\tau}$, $\Delta x_n=x_{n+1}-x_n$, and $F_n=\int_{t_n}^{t_{n+1}}f(x_s)\,\mathrm{d}s$.
Equation~\eqref{eq:exact} holds for any drift $f(x)$ and any ratio $\Delta t/\tau$.
The contribution of the hidden noise inherited from the previous interval cancels exactly, leaving only a stochastic term $\zeta_n$ that is an integral over the white noise increments in $[t_{n-1},t_{n+1}]$ (see \cite{Note1}).
Crucially, this implies that $\zeta_n$ is independent of all earlier positions $x_{n-1},x_{n-2},\ldots$, but correlated with $x_n$ and $x_{n+1}$.
\blue{Since $\tau$ is generally not known in advance, we infer it from the data together with the drift (see below).
In real-world applications such as power-grid frequency, the noise correlations are typically more complex than in Eq.~\eqref{eq:OU}.
In this case, the inferred $\rho=e^{-\Delta t/\tau}$ in Eq.~\eqref{eq:exact} acts as an effective noise correlation over one time step.
It still removes the dominant part of the correlations that KM estimation [Eq.~\eqref{eq:KM}] ignores entirely, which makes the approach more robust than KM estimation also beyond the noise model in Eq.~\eqref{eq:OU}.}

\blue{Equation~\eqref{eq:exact} represents a direct generalization of the KM estimator in Eq.~\eqref{eq:KM}, which is contained as the limit $\tau\to0$, i.e., $\rho\to 0$.
As we showed above, the KM estimator fails for any $\tau>0$, but we now see that this failure can be cured by introducing a $\rho$-weighted second increment.
To turn Eq.~\eqref{eq:exact} into a practical and robust estimator, however, several technical steps have to be performed.}

We approximate the drift integrals using the trapezoidal rule, $F_n\simeq\frac{\Delta t}{2}[f(x_n)+f(x_{n+1})]$.
This requires the sampling interval $\Delta t$ to be small enough to resolve the deterministic dynamics \blue{(exactly as in Eq.~\eqref{eq:KM})}, but imposes no restriction on $\Delta t/\tau$.
\blue{While the simple KM estimator \eqref{eq:KM} could be evaluated for each $x$ individually, Eq.~\eqref{eq:exact} contains $f(x_t)$ at different times.
However, this is not a problem since we can} expand the drift in a basis of functions, $f(x)=\sum_{k=1}^{K}\theta_k\phi_k(x)$, \blue{which turns} Eq.~\eqref{eq:exact} into a linear regression $y_n=\sum_{k=1}^K X_{nk} \theta_k + \zeta_n$ with regressors
\begin{equation}
  X_{nk}=\frac{\Delta t}{2}
  \left[
    \phi_k(x_{n+1})\!+\!(1-\rho)\phi_k(x_n)
    \!-\!\rho\phi_k(x_{n-1})
  \right].
\end{equation}
\blue{Each basis function $\phi_k$ covers a small interval of $x$ values (see~\cite{Note1} for implementation details), so the drift is resolved on the same finite spatial grid
as in standard KM estimation, where Eq.~\eqref{eq:KM} is evaluated in bins of
$x$~\cite{Gorjao2019}.}
There is, however, a complication: the regressors $X_{nk}$ depend on $x_n$ and $x_{n+1}$, which are correlated with the remaining noise $\zeta_n$.
Consequently, ordinary least squares yields systematically biased estimates, even for arbitrarily long trajectories. 
\blue{Intuitively, the remedy is to use only information from the past, which the remaining noise cannot have affected.}

This problem, known as \textit{endogeneity}, is well studied and can be overcome using \textit{instrumental variables}~\cite{Basmann1957,Sargan1958,Angrist2001,Hayashi2000} \blue{and \textit{moment conditions}~\cite{Hansen1982,HansenSingleton1982,Hall2010,Nikakhtar2023}.}
The idea is to replace the regressors in the estimation conditions by quantities that are informative about them but independent of the stochastic innovation.
The properties of $\zeta_n$ identified above provide a natural choice: functions of earlier positions $x_{n-1},x_{n-2},\ldots$.
Specifically, we choose the instruments $\phi_k(x_{n-1})$ and $\phi_k(x_{n-2})$ and impose the $2K$ moment conditions
\begin{equation}
  \left\langle
    \phi_k(x_{n-l})
    \left[y_n- \sum_{k'=1}^K X_{nk'} \theta_{k'} \right]
  \right\rangle=0 \,,
  \label{eq:moments}
\end{equation}
for $l=1,2$ and $k=1,\ldots, K$~\cite{Basmann1957,Sargan1958}.
Technically, this results in performing two linear regression steps instead of one to determine the coefficients $\btheta$ in $f(x)=\sum_{k=1}^{K}\theta_k\phi_k(x)$ \cite{Note1}.

So far, we have assumed that the noise correlation time $\tau$, and hence $\rho=e^{-\Delta t/\tau}$, is known. However, the same approach can also be used to infer $\tau$ from the observed trajectories.
For an incorrect value of $\rho$, the residuals in the regression retain part of the noise, which is correlated with earlier positions.
Consequently, the $2K$ moment conditions in Eq.~\eqref{eq:moments} generally cannot all be satisfied by adjusting the $K$ drift coefficients alone.
We therefore determine $\tau$ by minimizing the remaining mismatch using the generalized method of moments (GMM)~\cite{Hansen1982,HansenSingleton1982,Hall2010} (see \cite{Note1} for details).

\paragraph*{Applications.}
In Fig.~\ref{fig:fig2}, we evaluate the two-increment estimator in three examples, spanning numerical simulations and real-world dynamics in natural and engineered systems: a simulated asymmetric double well, the motion of a tracer in an active bacterial bath \cite{VillalobosConcha2025}, and frequency fluctuations in the GB power grid.
\blue{The examples provide increasingly demanding challenges: an exactly known drift, a linear drift predicted by an independently calibrated model, and a drift without a known ground truth, where the question is whether the method reveals physically meaningful structure that KM estimation misses.}
We compare our approach with the conventional KM estimator and its leading-order correction by the factor $2\tau/\Delta t$. \blue{
In all cases, $\tau$ is not an input but is inferred from the data: it is the value for which the remaining noise $\zeta_n$ in Eq.~\eqref{eq:exact} is least correlated with past positions, as quantified by the GMM criterion (details in~\cite{Note1}).
For the same data, the approach of Ref.~\cite{Willers2021} shows systematic deviations for the double well and the power grid (see~\cite{Note1}).}

For the asymmetric double well [\blue{Figs.~\ref{fig:fig2}(a) and \ref{fig:fig2}(b)}], the two-increment estimator accurately recovers the drift over the entire range, including the barrier region.
The GMM procedure yields a correlation time $\hat\tau=0.44$, close to the true value $\tau=0.40$.
\blue{The agreement with the ground truth of $\hat\tau$ and $\varphi(x)$ in Fig.~\ref{fig:fig2}(b) becomes almost exact when reducing $\Delta t$ from $0.1$ to $0.05$, see \cite{Note1}.} 
Importantly, the reconstructed potential $\varphi(x)$ reproduces both wells and the separating barrier.
In contrast, the KM-based estimates miss the shallow well entirely and substantially underestimate the depth of the deeper well\blue{, which would, e.g., grossly misestimate escape rates between the two states}.

We next consider experimental trajectories of a tracer droplet in an active bacterial bath~\cite{VillalobosConcha2025} [\blue{Fig.~\ref{fig:fig2}(c)}].
\blue{Here, the swimming bacteria generate the correlated noise, and the drift arises from the confinement of the oil droplet, which floats up against the curved wall of its enclosing droplet.}
The estimated correlation time $\hat\tau=0.21$\,s agrees with the bath correlation time $\tau_b=0.21$\,s obtained from an independently calibrated model~\cite{VillalobosConcha2025}. The reconstructed drift also
agrees well with the model prediction\blue{, whereas the conventional KM estimate is
several times too weak}. 

Finally, we analyze frequency fluctuations in the GB power grid \blue{[Fig.~\ref{fig:fig2}(d); data from \footnote{Power-grid frequency data for 2023 downloaded from NESO (National Energy System Operator, GB), \href{https://www.neso.energy/data-portal/system-frequency-data}{https://www.neso.energy/data-portal/system-frequency-data}, accessed 9 Sept 2026}].
Grid frequency reflects the balance between electricity generation and demand.
Imbalances shift the frequency away from its nominal value of 50\,Hz, and control mechanisms restore it, see Fig.~\ref{fig:fig1}(c).
Because these imbalances, e.g., from fluctuating demand, renewable generation, and electricity trading, are correlated over time, power-grid frequency is subject to strongly correlated noise.
Although these imbalances may involve multiple time scales, the GMM procedure (see \cite{Note1}) identifies a single effective correlation time $\hat\tau\approx5$\,min, which captures their dominant time scale \cite{gorbunov2022inertia}.}

The reconstructed drift reveals a pronounced nonlinear restoring force, best visible in the inset of Fig.~\ref{fig:fig2}(d). \blue{For
frequency deviations beyond $\pm20$\,mHz, the restoring rate (slope of the drift)  is about 0.36~$\mathrm{s}^{-1}$, corresponding to relaxation times of 2--3\,s, of the same order as literature values
of 1--2\,s for earlier years~\cite{Shinoda2023,Shouran2021}. 
Within $\pm15$\,mHz of the nominal frequency, the two-increment estimate gives a restoring rate that is smaller by a factor of about 3 [linear fits in the inset of Fig.~\ref{fig:fig2}(d)].}
The reduced restoring rate is consistent with deadbands in frequency control~\cite{vorobev2019deadbands}, demonstrating that the effective response of control mechanisms can be reconstructed directly from measured frequency trajectories despite correlated power imbalances.
\blue{Such deadbands in grid operation are introduced deliberately.
By responding less to small deviations,
generators save control effort and reduce mechanical wear.
However, frequency distributions measured in several power grids (including GB) are bimodal, with the frequency spending much of its time near the edges of the deadband, which in turn increases control activity and wear \cite{vorobev2019deadbands,delgiudice2021effects}.
How this bimodality arises is an open question; one explanation attributes it to the combination of a deadband-type nonlinear restoring force and correlated power imbalances~\cite{vorobev2019deadbands,delgiudice2021effects}.
Our estimator recovers both these ingredients directly from measured frequency data, which paves the way for data-driven approaches to this question.}

In contrast, the conventional KM estimator [Eq.~\eqref{eq:KM}, gray in Fig.~\ref{fig:fig2}(d)] underestimates the restoring force by \blue{more than two orders of magnitude, as expected from $\mathrm{KM}(\Delta t)\approx(\Delta t/2\tau)f$, see Eq.~\eqref{eq:km_first}}.
\blue{The corrected KM estimate recovers an outer restoring force of similar magnitude, but near the nominal frequency it scatters strongly and even yields} a locally increasing drift $f(x)$, suggesting a spurious destabilizing
force.

\paragraph*{Discussion.}
\blue{For decades, Kramers--Moyal estimation has been central for extracting forces from fluctuating trajectories, valued for its simplicity and its independence from detailed models, and it has shaped the analysis of systems from turbulence and financial markets to climate records and power grids.
Yet an increasing number of contemporary challenges feature correlated noise, which} does not merely add noise to this inference---it can hide the forces entirely.
\blue{Whenever unobserved degrees of freedom carry memory, averages conditioned on the present state mix the deterministic force with the fluctuations that brought the system there, obscuring the inference.}
\blue{Since this failure follows from stationarity and
differentiable trajectories alone, independently of the detailed noise statistics, it
concerns a broad class of stochastic systems.}

\blue{Conditioning on the past instead of the present, alongside using two increments, resolves this problem while retaining what made KM estimation so successful: the inference remains direct, fast, and interpretable.
Although our estimator is exact only for OU noise, its result for grid frequency, where the noise involves many time scales, suggests that a single effective correlation time already captures much of what classical KM analysis misses.
This opens the way to revisiting the many systems in which KM methods have been applied despite noise correlations, and to systems previously out of reach, e.g., in active matter~\cite{Fodor2016,Martin2021,Maggi2014}, power grids with correlation effects increasing due to renewable generation~\cite{milano2013systematic,Milan2013,Anvari2016Short,tyloo2021primary},
and climate variability~\cite{Hasselmann1976,frankignoul1985sea,Morr2024,Morr2025,Riechers2025} on time scales comparable to their noise correlation.}
Natural extensions of the approach include noise with several correlation times, multiplicative noise, diffusion estimates and even higher KM moments, as well as higher dimensions. 

\blue{\paragraph*{Data availability.} The power-grid frequency data are publicly available from
NESO~\cite{Note2}. The bacterial-bath trajectories are published alongside 
Ref.~\cite{VillalobosConcha2025} at \cite{VillalobosConcha2025Zenodo}. The code used to produce Fig.~\ref{fig:fig2} will be made openly available on Zenodo upon publication.}

\paragraph*{Acknowledgements.}
CD acknowledges funding through the Walter Benjamin program by the Deutsche Forschungsgemeinschaft (DFG, German Research Foundation, project No. 570237743).
LRG acknowledges funding through the NNLL project by NordForsk (NordForsk, project No. 242039).

\bibliographystyle{apsrev4-2}
\bibliography{references.bib}

\begin{thebibliography}{86}%
\makeatletter
\providecommand \@ifxundefined [1]{%
 \@ifx{#1\undefined}
}%
\providecommand \@ifnum [1]{%
 \ifnum #1\expandafter \@firstoftwo
 \else \expandafter \@secondoftwo
 \fi
}%
\providecommand \@ifx [1]{%
 \ifx #1\expandafter \@firstoftwo
 \else \expandafter \@secondoftwo
 \fi
}%
\providecommand \natexlab [1]{#1}%
\providecommand \enquote  [1]{``#1''}%
\providecommand \bibnamefont  [1]{#1}%
\providecommand \bibfnamefont [1]{#1}%
\providecommand \citenamefont [1]{#1}%
\providecommand \href@noop [0]{\@secondoftwo}%
\providecommand \href [0]{\begingroup \@sanitize@url \@href}%
\providecommand \@href[1]{\@@startlink{#1}\@@href}%
\providecommand \@@href[1]{\endgroup#1\@@endlink}%
\providecommand \@sanitize@url [0]{\catcode `\\12\catcode `\$12\catcode
  `\&12\catcode `\#12\catcode `\^12\catcode `\_12\catcode `\%12\relax}%
\providecommand \@@startlink[1]{}%
\providecommand \@@endlink[0]{}%
\providecommand \url  [0]{\begingroup\@sanitize@url \@url }%
\providecommand \@url [1]{\endgroup\@href {#1}{\urlprefix }}%
\providecommand \urlprefix  [0]{URL }%
\providecommand \Eprint [0]{\href }%
\providecommand \doibase [0]{https://doi.org/}%
\providecommand \selectlanguage [0]{\@gobble}%
\providecommand \bibinfo  [0]{\@secondoftwo}%
\providecommand \bibfield  [0]{\@secondoftwo}%
\providecommand \translation [1]{[#1]}%
\providecommand \BibitemOpen [0]{}%
\providecommand \bibitemStop [0]{}%
\providecommand \bibitemNoStop [0]{.\EOS\space}%
\providecommand \EOS [0]{\spacefactor3000\relax}%
\providecommand \BibitemShut  [1]{\csname bibitem#1\endcsname}%
\let\auto@bib@innerbib\@empty
\bibitem [{\citenamefont {Friedrich}\ \emph {et~al.}(2011)\citenamefont
  {Friedrich}, \citenamefont {Peinke}, \citenamefont {Sahimi},\ and\
  \citenamefont {Reza Rahimi~Tabar}}]{friedrich2011approaching}%
  \BibitemOpen
  \bibfield  {author} {\bibinfo {author} {\bibfnamefont {R.}~\bibnamefont
  {Friedrich}}, \bibinfo {author} {\bibfnamefont {J.}~\bibnamefont {Peinke}},
  \bibinfo {author} {\bibfnamefont {M.}~\bibnamefont {Sahimi}},\ and\ \bibinfo
  {author} {\bibfnamefont {M.}~\bibnamefont {Reza Rahimi~Tabar}},\ }\href
  {https://doi.org/10.1016/j.physrep.2011.05.003} {\bibfield  {journal}
  {\bibinfo  {journal} {Physics Reports}\ }\textbf {\bibinfo {volume} {506}},\
  \bibinfo {pages} {87–162} (\bibinfo {year} {2011})}\BibitemShut {NoStop}%
\bibitem [{\citenamefont {Peinke}\ \emph {et~al.}(2019)\citenamefont {Peinke},
  \citenamefont {Tabar},\ and\ \citenamefont {W\"{a}chter}}]{Peinke2019}%
  \BibitemOpen
  \bibfield  {author} {\bibinfo {author} {\bibfnamefont {J.}~\bibnamefont
  {Peinke}}, \bibinfo {author} {\bibfnamefont {M.}~\bibnamefont {Tabar}},\ and\
  \bibinfo {author} {\bibfnamefont {M.}~\bibnamefont {W\"{a}chter}},\ }\href
  {https://doi.org/10.1146/annurev-conmatphys-033117-054252} {\bibfield
  {journal} {\bibinfo  {journal} {Annual Review of Condensed Matter Physics}\
  }\textbf {\bibinfo {volume} {10}},\ \bibinfo {pages} {107–132} (\bibinfo
  {year} {2019})}\BibitemShut {NoStop}%
\bibitem [{\citenamefont {Rahimi~Tabar}(2019)}]{RahimiTabar2019}%
  \BibitemOpen
  \bibfield  {author} {\bibinfo {author} {\bibfnamefont {M.~R.}\ \bibnamefont
  {Rahimi~Tabar}},\ }\href {https://doi.org/10.1007/978-3-030-18472-8} {\emph
  {\bibinfo {title} {Analysis and Data-Based Reconstruction of Complex
  Nonlinear Dynamical Systems: Using the Methods of Stochastic Processes}}}\
  (\bibinfo  {publisher} {Springer International Publishing},\ \bibinfo {year}
  {2019})\BibitemShut {NoStop}%
\bibitem [{\citenamefont {Siegert}\ \emph {et~al.}(1998)\citenamefont
  {Siegert}, \citenamefont {Friedrich},\ and\ \citenamefont
  {Peinke}}]{siegert1998analysis}%
  \BibitemOpen
  \bibfield  {author} {\bibinfo {author} {\bibfnamefont {S.}~\bibnamefont
  {Siegert}}, \bibinfo {author} {\bibfnamefont {R.}~\bibnamefont {Friedrich}},\
  and\ \bibinfo {author} {\bibfnamefont {J.}~\bibnamefont {Peinke}},\ }\href
  {https://doi.org/10.1016/s0375-9601(98)00283-7} {\bibfield  {journal}
  {\bibinfo  {journal} {Physics Letters A}\ }\textbf {\bibinfo {volume}
  {243}},\ \bibinfo {pages} {275–280} (\bibinfo {year} {1998})}\BibitemShut
  {NoStop}%
\bibitem [{\citenamefont {Gardiner}(2004)}]{gardiner2004handbook}%
  \BibitemOpen
  \bibfield  {author} {\bibinfo {author} {\bibfnamefont {C.~W.}\ \bibnamefont
  {Gardiner}},\ }\href@noop {} {\emph {\bibinfo {title} {Handbook of stochastic
  methods for physics, chemistry and the natural sciences}}},\ \bibinfo
  {edition} {3rd}\ ed.,\ \bibinfo {series} {Springer Series in Synergetics},
  Vol.~\bibinfo {volume} {13}\ (\bibinfo  {publisher} {Springer-Verlag},\
  \bibinfo {address} {Berlin},\ \bibinfo {year} {2004})\ pp.\ \bibinfo {pages}
  {xviii+415}\BibitemShut {NoStop}%
\bibitem [{\citenamefont {Øksendal}(2003)}]{oksendal2003stochastic}%
  \BibitemOpen
  \bibfield  {author} {\bibinfo {author} {\bibfnamefont {B.}~\bibnamefont
  {Øksendal}},\ }\href {https://doi.org/10.1007/978-3-642-14394-6} {\emph
  {\bibinfo {title} {Stochastic Differential Equations}}}\ (\bibinfo
  {publisher} {Springer Berlin Heidelberg},\ \bibinfo {year}
  {2003})\BibitemShut {NoStop}%
\bibitem [{\citenamefont {Friedrich}\ and\ \citenamefont
  {Peinke}(1997)}]{Friedrich1997}%
  \BibitemOpen
  \bibfield  {author} {\bibinfo {author} {\bibfnamefont {R.}~\bibnamefont
  {Friedrich}}\ and\ \bibinfo {author} {\bibfnamefont {J.}~\bibnamefont
  {Peinke}},\ }\href {https://doi.org/10.1103/physrevlett.78.863} {\bibfield
  {journal} {\bibinfo  {journal} {Physical Review Letters}\ }\textbf {\bibinfo
  {volume} {78}},\ \bibinfo {pages} {863–866} (\bibinfo {year}
  {1997})}\BibitemShut {NoStop}%
\bibitem [{\citenamefont {Friedrich}\ and\ \citenamefont
  {Grauer}(2020)}]{Friedrich2020}%
  \BibitemOpen
  \bibfield  {author} {\bibinfo {author} {\bibfnamefont {J.}~\bibnamefont
  {Friedrich}}\ and\ \bibinfo {author} {\bibfnamefont {R.}~\bibnamefont
  {Grauer}},\ }\href {https://doi.org/10.3390/atmos11091003} {\bibfield
  {journal} {\bibinfo  {journal} {Atmosphere}\ }\textbf {\bibinfo {volume}
  {11}},\ \bibinfo {pages} {1003} (\bibinfo {year} {2020})}\BibitemShut
  {NoStop}%
\bibitem [{\citenamefont {Friedrich}\ \emph {et~al.}(2000)\citenamefont
  {Friedrich}, \citenamefont {Peinke},\ and\ \citenamefont
  {Renner}}]{Friedrich2000}%
  \BibitemOpen
  \bibfield  {author} {\bibinfo {author} {\bibfnamefont {R.}~\bibnamefont
  {Friedrich}}, \bibinfo {author} {\bibfnamefont {J.}~\bibnamefont {Peinke}},\
  and\ \bibinfo {author} {\bibfnamefont {C.}~\bibnamefont {Renner}},\ }\href
  {https://doi.org/10.1103/physrevlett.84.5224} {\bibfield  {journal} {\bibinfo
   {journal} {Physical Review Letters}\ }\textbf {\bibinfo {volume} {84}},\
  \bibinfo {pages} {5224–5227} (\bibinfo {year} {2000})}\BibitemShut
  {NoStop}%
\bibitem [{\citenamefont {Br\"{u}ckner}\ \emph {et~al.}(2019)\citenamefont
  {Br\"{u}ckner}, \citenamefont {Fink}, \citenamefont {Schreiber},
  \citenamefont {R\"{o}ttgermann}, \citenamefont {R\"{a}dler},\ and\
  \citenamefont {Broedersz}}]{Brueckner2019}%
  \BibitemOpen
  \bibfield  {author} {\bibinfo {author} {\bibfnamefont {D.~B.}\ \bibnamefont
  {Br\"{u}ckner}}, \bibinfo {author} {\bibfnamefont {A.}~\bibnamefont {Fink}},
  \bibinfo {author} {\bibfnamefont {C.}~\bibnamefont {Schreiber}}, \bibinfo
  {author} {\bibfnamefont {P.~J.~F.}\ \bibnamefont {R\"{o}ttgermann}}, \bibinfo
  {author} {\bibfnamefont {J.~O.}\ \bibnamefont {R\"{a}dler}},\ and\ \bibinfo
  {author} {\bibfnamefont {C.~P.}\ \bibnamefont {Broedersz}},\ }\href
  {https://doi.org/10.1038/s41567-019-0445-4} {\bibfield  {journal} {\bibinfo
  {journal} {Nat. Phys.}\ }\textbf {\bibinfo {volume} {15}},\ \bibinfo {pages}
  {595–601} (\bibinfo {year} {2019})}\BibitemShut {NoStop}%
\bibitem [{\citenamefont {Prusseit}\ and\ \citenamefont
  {Lehnertz}(2007)}]{Prusseit2007}%
  \BibitemOpen
  \bibfield  {author} {\bibinfo {author} {\bibfnamefont {J.}~\bibnamefont
  {Prusseit}}\ and\ \bibinfo {author} {\bibfnamefont {K.}~\bibnamefont
  {Lehnertz}},\ }\href {https://doi.org/10.1103/physrevlett.98.138103}
  {\bibfield  {journal} {\bibinfo  {journal} {Physical Review Letters}\
  }\textbf {\bibinfo {volume} {98}},\ \bibinfo {pages} {138103} (\bibinfo
  {year} {2007})}\BibitemShut {NoStop}%
\bibitem [{\citenamefont {Lamouroux}\ and\ \citenamefont
  {Lehnertz}(2009)}]{Lamouroux2009}%
  \BibitemOpen
  \bibfield  {author} {\bibinfo {author} {\bibfnamefont {D.}~\bibnamefont
  {Lamouroux}}\ and\ \bibinfo {author} {\bibfnamefont {K.}~\bibnamefont
  {Lehnertz}},\ }\href {https://doi.org/10.1016/j.physleta.2009.07.073}
  {\bibfield  {journal} {\bibinfo  {journal} {Physics Letters A}\ }\textbf
  {\bibinfo {volume} {373}},\ \bibinfo {pages} {3507} (\bibinfo {year}
  {2009})}\BibitemShut {NoStop}%
\bibitem [{\citenamefont {Hashtroud}\ \emph {et~al.}(2019)\citenamefont
  {Hashtroud}, \citenamefont {Mirzahossein}, \citenamefont {Zarei},\ and\
  \citenamefont {Tabar}}]{Hashtroud2019}%
  \BibitemOpen
  \bibfield  {author} {\bibinfo {author} {\bibfnamefont {A.~M.}\ \bibnamefont
  {Hashtroud}}, \bibinfo {author} {\bibfnamefont {E.}~\bibnamefont
  {Mirzahossein}}, \bibinfo {author} {\bibfnamefont {F.}~\bibnamefont
  {Zarei}},\ and\ \bibinfo {author} {\bibfnamefont {M.~R.~R.}\ \bibnamefont
  {Tabar}},\ }\href {https://doi.org/10.1088/1742-5468/ab3287} {\bibfield
  {journal} {\bibinfo  {journal} {Journal of Statistical Mechanics: Theory and
  Experiment}\ }\textbf {\bibinfo {volume} {2019}},\ \bibinfo {pages} {083213}
  (\bibinfo {year} {2019})}\BibitemShut {NoStop}%
\bibitem [{\citenamefont {Ditlevsen}(1999)}]{Ditlevsen1999}%
  \BibitemOpen
  \bibfield  {author} {\bibinfo {author} {\bibfnamefont {P.~D.}\ \bibnamefont
  {Ditlevsen}},\ }\href {https://doi.org/10.1029/1999gl900252} {\bibfield
  {journal} {\bibinfo  {journal} {Geophysical Research Letters}\ }\textbf
  {\bibinfo {volume} {26}},\ \bibinfo {pages} {1441–1444} (\bibinfo {year}
  {1999})}\BibitemShut {NoStop}%
\bibitem [{\citenamefont {Riechers}\ \emph {et~al.}(2025)\citenamefont
  {Riechers}, \citenamefont {Morr}, \citenamefont {Lehnertz}, \citenamefont
  {Lind}, \citenamefont {Boers}, \citenamefont {Witthaut},\ and\ \citenamefont
  {Rydin Gorj\~ao}}]{Riechers2025}%
  \BibitemOpen
  \bibfield  {author} {\bibinfo {author} {\bibfnamefont {K.}~\bibnamefont
  {Riechers}}, \bibinfo {author} {\bibfnamefont {A.}~\bibnamefont {Morr}},
  \bibinfo {author} {\bibfnamefont {K.}~\bibnamefont {Lehnertz}}, \bibinfo
  {author} {\bibfnamefont {P.~G.}\ \bibnamefont {Lind}}, \bibinfo {author}
  {\bibfnamefont {N.}~\bibnamefont {Boers}}, \bibinfo {author} {\bibfnamefont
  {D.}~\bibnamefont {Witthaut}},\ and\ \bibinfo {author} {\bibfnamefont
  {L.}~\bibnamefont {Rydin Gorj\~ao}},\ }\href
  {https://doi.org/10.1007/s00382-025-07880-9} {\bibfield  {journal} {\bibinfo
  {journal} {Climate Dynamics}\ }\textbf {\bibinfo {volume} {63}},\ \bibinfo
  {pages} {465} (\bibinfo {year} {2025})}\BibitemShut {NoStop}%
\bibitem [{\citenamefont {Milan}\ \emph {et~al.}(2013)\citenamefont {Milan},
  \citenamefont {W\"{a}chter},\ and\ \citenamefont {Peinke}}]{Milan2013}%
  \BibitemOpen
  \bibfield  {author} {\bibinfo {author} {\bibfnamefont {P.}~\bibnamefont
  {Milan}}, \bibinfo {author} {\bibfnamefont {M.}~\bibnamefont {W\"{a}chter}},\
  and\ \bibinfo {author} {\bibfnamefont {J.}~\bibnamefont {Peinke}},\ }\href
  {https://doi.org/10.1103/physrevlett.110.138701} {\bibfield  {journal}
  {\bibinfo  {journal} {Physical Review Letters}\ }\textbf {\bibinfo {volume}
  {110}},\ \bibinfo {pages} {138701} (\bibinfo {year} {2013})}\BibitemShut
  {NoStop}%
\bibitem [{\citenamefont {Rydin~Gorj{\~a}o}\ \emph {et~al.}(2020)\citenamefont
  {Rydin~Gorj{\~a}o}, \citenamefont {Anvari}, \citenamefont {Kantz},
  \citenamefont {Beck}, \citenamefont {Witthaut}, \citenamefont {Timme},\ and\
  \citenamefont {Sch{\"a}fer}}]{rydin2020data}%
  \BibitemOpen
  \bibfield  {author} {\bibinfo {author} {\bibfnamefont {L.}~\bibnamefont
  {Rydin~Gorj{\~a}o}}, \bibinfo {author} {\bibfnamefont {M.}~\bibnamefont
  {Anvari}}, \bibinfo {author} {\bibfnamefont {H.}~\bibnamefont {Kantz}},
  \bibinfo {author} {\bibfnamefont {C.}~\bibnamefont {Beck}}, \bibinfo {author}
  {\bibfnamefont {D.}~\bibnamefont {Witthaut}}, \bibinfo {author}
  {\bibfnamefont {M.}~\bibnamefont {Timme}},\ and\ \bibinfo {author}
  {\bibfnamefont {B.}~\bibnamefont {Sch{\"a}fer}},\ }\href
  {https://doi.org/10.1109/access.2020.2967834} {\bibfield  {journal} {\bibinfo
   {journal} {IEEE access}\ }\textbf {\bibinfo {volume} {8}},\ \bibinfo {pages}
  {43082} (\bibinfo {year} {2020})}\BibitemShut {NoStop}%
\bibitem [{\citenamefont {Rydin Gorj\~ao}\ \emph {et~al.}(2020)\citenamefont
  {Rydin Gorj\~ao}, \citenamefont {Jumar}, \citenamefont {Maass}, \citenamefont
  {Hagenmeyer}, \citenamefont {Yalcin}, \citenamefont {Kruse}, \citenamefont
  {Timme}, \citenamefont {Beck}, \citenamefont {Witthaut},\ and\ \citenamefont
  {Sch\"{a}fer}}]{RydinGorjao2020}%
  \BibitemOpen
  \bibfield  {author} {\bibinfo {author} {\bibfnamefont {L.}~\bibnamefont
  {Rydin Gorj\~ao}}, \bibinfo {author} {\bibfnamefont {R.}~\bibnamefont
  {Jumar}}, \bibinfo {author} {\bibfnamefont {H.}~\bibnamefont {Maass}},
  \bibinfo {author} {\bibfnamefont {V.}~\bibnamefont {Hagenmeyer}}, \bibinfo
  {author} {\bibfnamefont {G.~C.}\ \bibnamefont {Yalcin}}, \bibinfo {author}
  {\bibfnamefont {J.}~\bibnamefont {Kruse}}, \bibinfo {author} {\bibfnamefont
  {M.}~\bibnamefont {Timme}}, \bibinfo {author} {\bibfnamefont
  {C.}~\bibnamefont {Beck}}, \bibinfo {author} {\bibfnamefont {D.}~\bibnamefont
  {Witthaut}},\ and\ \bibinfo {author} {\bibfnamefont {B.}~\bibnamefont
  {Sch\"{a}fer}},\ }\href {https://doi.org/10.1038/s41467-020-19732-7}
  {\bibfield  {journal} {\bibinfo  {journal} {Nature Communications}\ }\textbf
  {\bibinfo {volume} {11}},\ \bibinfo {pages} {6362} (\bibinfo {year}
  {2020})}\BibitemShut {NoStop}%
\bibitem [{\citenamefont {Stanton}(1997)}]{Stanton1997}%
  \BibitemOpen
  \bibfield  {author} {\bibinfo {author} {\bibfnamefont {R.}~\bibnamefont
  {Stanton}},\ }\href {https://doi.org/10.1111/j.1540-6261.1997.tb02748.x}
  {\bibfield  {journal} {\bibinfo  {journal} {The Journal of Finance}\ }\textbf
  {\bibinfo {volume} {52}},\ \bibinfo {pages} {1973–2002} (\bibinfo {year}
  {1997})}\BibitemShut {NoStop}%
\bibitem [{\citenamefont {Aït-Sahalia}(1996)}]{AtSahalia1996}%
  \BibitemOpen
  \bibfield  {author} {\bibinfo {author} {\bibfnamefont {Y.}~\bibnamefont
  {Aït-Sahalia}},\ }\href {https://doi.org/10.1093/rfs/9.2.385} {\bibfield
  {journal} {\bibinfo  {journal} {Review of Financial Studies}\ }\textbf
  {\bibinfo {volume} {9}},\ \bibinfo {pages} {385–426} (\bibinfo {year}
  {1996})}\BibitemShut {NoStop}%
\bibitem [{\citenamefont {Bandi}\ and\ \citenamefont
  {Phillips}(2003)}]{Bandi2003}%
  \BibitemOpen
  \bibfield  {author} {\bibinfo {author} {\bibfnamefont {F.~M.}\ \bibnamefont
  {Bandi}}\ and\ \bibinfo {author} {\bibfnamefont {P.~C.~B.}\ \bibnamefont
  {Phillips}},\ }\href {https://doi.org/10.1111/1468-0262.00395} {\bibfield
  {journal} {\bibinfo  {journal} {Econometrica}\ }\textbf {\bibinfo {volume}
  {71}},\ \bibinfo {pages} {241–283} (\bibinfo {year} {2003})}\BibitemShut
  {NoStop}%
\bibitem [{\citenamefont {Frishman}\ and\ \citenamefont
  {Ronceray}(2020)}]{Frishman2020}%
  \BibitemOpen
  \bibfield  {author} {\bibinfo {author} {\bibfnamefont {A.}~\bibnamefont
  {Frishman}}\ and\ \bibinfo {author} {\bibfnamefont {P.}~\bibnamefont
  {Ronceray}},\ }\href {https://doi.org/10.1103/physrevx.10.021009} {\bibfield
  {journal} {\bibinfo  {journal} {Physical Review X}\ }\textbf {\bibinfo
  {volume} {10}},\ \bibinfo {pages} {021009} (\bibinfo {year}
  {2020})}\BibitemShut {NoStop}%
\bibitem [{\citenamefont {Ferretti}\ \emph {et~al.}(2020)\citenamefont
  {Ferretti}, \citenamefont {Chardès}, \citenamefont {Mora}, \citenamefont
  {Walczak},\ and\ \citenamefont {Giardina}}]{ferretti2020building}%
  \BibitemOpen
  \bibfield  {author} {\bibinfo {author} {\bibfnamefont {F.}~\bibnamefont
  {Ferretti}}, \bibinfo {author} {\bibfnamefont {V.}~\bibnamefont {Chardès}},
  \bibinfo {author} {\bibfnamefont {T.}~\bibnamefont {Mora}}, \bibinfo {author}
  {\bibfnamefont {A.~M.}\ \bibnamefont {Walczak}},\ and\ \bibinfo {author}
  {\bibfnamefont {I.}~\bibnamefont {Giardina}},\ }\href
  {https://doi.org/10.1103/physrevx.10.031018} {\bibfield  {journal} {\bibinfo
  {journal} {Physical Review X}\ }\textbf {\bibinfo {volume} {10}},\ \bibinfo
  {pages} {031018} (\bibinfo {year} {2020})}\BibitemShut {NoStop}%
\bibitem [{\citenamefont {Br\"{u}ckner}\ \emph {et~al.}(2020)\citenamefont
  {Br\"{u}ckner}, \citenamefont {Ronceray},\ and\ \citenamefont
  {Broedersz}}]{Brueckner2020}%
  \BibitemOpen
  \bibfield  {author} {\bibinfo {author} {\bibfnamefont {D.~B.}\ \bibnamefont
  {Br\"{u}ckner}}, \bibinfo {author} {\bibfnamefont {P.}~\bibnamefont
  {Ronceray}},\ and\ \bibinfo {author} {\bibfnamefont {C.~P.}\ \bibnamefont
  {Broedersz}},\ }\href {https://doi.org/10.1103/physrevlett.125.058103}
  {\bibfield  {journal} {\bibinfo  {journal} {Physical Review Letters}\
  }\textbf {\bibinfo {volume} {125}},\ \bibinfo {pages} {058103} (\bibinfo
  {year} {2020})}\BibitemShut {NoStop}%
\bibitem [{\citenamefont {Zwanzig}(1961)}]{Zwanzig1961}%
  \BibitemOpen
  \bibfield  {author} {\bibinfo {author} {\bibfnamefont {R.}~\bibnamefont
  {Zwanzig}},\ }\href {https://doi.org/10.1103/physrev.124.983} {\bibfield
  {journal} {\bibinfo  {journal} {Physical Review}\ }\textbf {\bibinfo {volume}
  {124}},\ \bibinfo {pages} {983–992} (\bibinfo {year} {1961})}\BibitemShut
  {NoStop}%
\bibitem [{\citenamefont {Mori}(1965)}]{Mori1965}%
  \BibitemOpen
  \bibfield  {author} {\bibinfo {author} {\bibfnamefont {H.}~\bibnamefont
  {Mori}},\ }\href {https://doi.org/10.1143/ptp.33.423} {\bibfield  {journal}
  {\bibinfo  {journal} {Progress of Theoretical Physics}\ }\textbf {\bibinfo
  {volume} {33}},\ \bibinfo {pages} {423–455} (\bibinfo {year}
  {1965})}\BibitemShut {NoStop}%
\bibitem [{\citenamefont {Wu}\ and\ \citenamefont {Libchaber}(2000)}]{Wu2000}%
  \BibitemOpen
  \bibfield  {author} {\bibinfo {author} {\bibfnamefont {X.-L.}\ \bibnamefont
  {Wu}}\ and\ \bibinfo {author} {\bibfnamefont {A.}~\bibnamefont {Libchaber}},\
  }\href {https://doi.org/10.1103/physrevlett.84.3017} {\bibfield  {journal}
  {\bibinfo  {journal} {Physical Review Letters}\ }\textbf {\bibinfo {volume}
  {84}},\ \bibinfo {pages} {3017–3020} (\bibinfo {year} {2000})}\BibitemShut
  {NoStop}%
\bibitem [{\citenamefont {Maggi}\ \emph {et~al.}(2014)\citenamefont {Maggi},
  \citenamefont {Paoluzzi}, \citenamefont {Pellicciotta}, \citenamefont
  {Lepore}, \citenamefont {Angelani},\ and\ \citenamefont
  {Di~Leonardo}}]{Maggi2014}%
  \BibitemOpen
  \bibfield  {author} {\bibinfo {author} {\bibfnamefont {C.}~\bibnamefont
  {Maggi}}, \bibinfo {author} {\bibfnamefont {M.}~\bibnamefont {Paoluzzi}},
  \bibinfo {author} {\bibfnamefont {N.}~\bibnamefont {Pellicciotta}}, \bibinfo
  {author} {\bibfnamefont {A.}~\bibnamefont {Lepore}}, \bibinfo {author}
  {\bibfnamefont {L.}~\bibnamefont {Angelani}},\ and\ \bibinfo {author}
  {\bibfnamefont {R.}~\bibnamefont {Di~Leonardo}},\ }\href
  {https://doi.org/10.1103/physrevlett.113.238303} {\bibfield  {journal}
  {\bibinfo  {journal} {Physical Review Letters}\ }\textbf {\bibinfo {volume}
  {113}},\ \bibinfo {pages} {238303} (\bibinfo {year} {2014})}\BibitemShut
  {NoStop}%
\bibitem [{\citenamefont {Villalobos-Concha}\ \emph {et~al.}(2025)\citenamefont
  {Villalobos-Concha}, \citenamefont {Liu}, \citenamefont {Ramos},
  \citenamefont {Goral}, \citenamefont {Lindner}, \citenamefont
  {L\'opez-Le\'on}, \citenamefont {Clément}, \citenamefont {Soto},\ and\
  \citenamefont {Cordero}}]{VillalobosConcha2025}%
  \BibitemOpen
  \bibfield  {author} {\bibinfo {author} {\bibfnamefont {C.}~\bibnamefont
  {Villalobos-Concha}}, \bibinfo {author} {\bibfnamefont {Z.}~\bibnamefont
  {Liu}}, \bibinfo {author} {\bibfnamefont {G.}~\bibnamefont {Ramos}}, \bibinfo
  {author} {\bibfnamefont {M.}~\bibnamefont {Goral}}, \bibinfo {author}
  {\bibfnamefont {A.}~\bibnamefont {Lindner}}, \bibinfo {author} {\bibfnamefont
  {T.}~\bibnamefont {L\'opez-Le\'on}}, \bibinfo {author} {\bibfnamefont
  {E.}~\bibnamefont {Clément}}, \bibinfo {author} {\bibfnamefont
  {R.}~\bibnamefont {Soto}},\ and\ \bibinfo {author} {\bibfnamefont {M.~L.}\
  \bibnamefont {Cordero}},\ }\href {https://doi.org/10.1073/pnas.2426096122}
  {\bibfield  {journal} {\bibinfo  {journal} {Proc. Natl. Acad. Sci. U.S.A.}\
  }\textbf {\bibinfo {volume} {122}},\ \bibinfo {pages} {e2426096122} (\bibinfo
  {year} {2025})}\BibitemShut {NoStop}%
\bibitem [{\citenamefont {Ginot}\ \emph {et~al.}(2022)\citenamefont {Ginot},
  \citenamefont {Caspers}, \citenamefont {Kr{\"u}ger},\ and\ \citenamefont
  {Bechinger}}]{Ginot2022}%
  \BibitemOpen
  \bibfield  {author} {\bibinfo {author} {\bibfnamefont {F.}~\bibnamefont
  {Ginot}}, \bibinfo {author} {\bibfnamefont {J.}~\bibnamefont {Caspers}},
  \bibinfo {author} {\bibfnamefont {M.}~\bibnamefont {Kr{\"u}ger}},\ and\
  \bibinfo {author} {\bibfnamefont {C.}~\bibnamefont {Bechinger}},\ }\href
  {https://doi.org/10.1103/PhysRevLett.128.028001} {\bibfield  {journal}
  {\bibinfo  {journal} {Physical Review Letters}\ }\textbf {\bibinfo {volume}
  {128}},\ \bibinfo {pages} {028001} (\bibinfo {year} {2022})}\BibitemShut
  {NoStop}%
\bibitem [{\citenamefont {Hasselmann}(1976)}]{Hasselmann1976}%
  \BibitemOpen
  \bibfield  {author} {\bibinfo {author} {\bibfnamefont {K.}~\bibnamefont
  {Hasselmann}},\ }\href {https://doi.org/10.1111/j.2153-3490.1976.tb00696.x}
  {\bibfield  {journal} {\bibinfo  {journal} {Tellus}\ }\textbf {\bibinfo
  {volume} {28}},\ \bibinfo {pages} {473–485} (\bibinfo {year}
  {1976})}\BibitemShut {NoStop}%
\bibitem [{\citenamefont {Frankignoul}(1985)}]{frankignoul1985sea}%
  \BibitemOpen
  \bibfield  {author} {\bibinfo {author} {\bibfnamefont {C.}~\bibnamefont
  {Frankignoul}},\ }\href {https://doi.org/10.1029/rg023i004p00357} {\bibfield
  {journal} {\bibinfo  {journal} {Reviews of Geophysics}\ }\textbf {\bibinfo
  {volume} {23}},\ \bibinfo {pages} {357–390} (\bibinfo {year}
  {1985})}\BibitemShut {NoStop}%
\bibitem [{\citenamefont {Morr}\ and\ \citenamefont {Boers}(2024)}]{Morr2024}%
  \BibitemOpen
  \bibfield  {author} {\bibinfo {author} {\bibfnamefont {A.}~\bibnamefont
  {Morr}}\ and\ \bibinfo {author} {\bibfnamefont {N.}~\bibnamefont {Boers}},\
  }\href {https://doi.org/10.1103/physrevx.14.021037} {\bibfield  {journal}
  {\bibinfo  {journal} {Physical Review X}\ }\textbf {\bibinfo {volume} {14}},\
  \bibinfo {pages} {021037} (\bibinfo {year} {2024})}\BibitemShut {NoStop}%
\bibitem [{\citenamefont {Morr}\ \emph {et~al.}(2025)\citenamefont {Morr},
  \citenamefont {Kreher},\ and\ \citenamefont {Boers}}]{Morr2025}%
  \BibitemOpen
  \bibfield  {author} {\bibinfo {author} {\bibfnamefont {A.}~\bibnamefont
  {Morr}}, \bibinfo {author} {\bibfnamefont {D.}~\bibnamefont {Kreher}},\ and\
  \bibinfo {author} {\bibfnamefont {N.}~\bibnamefont {Boers}},\ }\href
  {https://doi.org/10.1098/rsos.250573} {\bibfield  {journal} {\bibinfo
  {journal} {Royal Society Open Science}\ }\textbf {\bibinfo {volume} {12}},\
  \bibinfo {pages} {250573} (\bibinfo {year} {2025})}\BibitemShut {NoStop}%
\bibitem [{\citenamefont {Milano}\ and\ \citenamefont
  {Zarate-Minano}(2013)}]{milano2013systematic}%
  \BibitemOpen
  \bibfield  {author} {\bibinfo {author} {\bibfnamefont {F.}~\bibnamefont
  {Milano}}\ and\ \bibinfo {author} {\bibfnamefont {R.}~\bibnamefont
  {Zarate-Minano}},\ }\href {https://doi.org/10.1109/tpwrs.2013.2266441}
  {\bibfield  {journal} {\bibinfo  {journal} {IEEE Transactions on Power
  Systems}\ }\textbf {\bibinfo {volume} {28}},\ \bibinfo {pages} {4537–4544}
  (\bibinfo {year} {2013})}\BibitemShut {NoStop}%
\bibitem [{\citenamefont {Roberts}\ \emph {et~al.}(2016)\citenamefont
  {Roberts}, \citenamefont {Stewart},\ and\ \citenamefont
  {Milano}}]{roberts2016validation}%
  \BibitemOpen
  \bibfield  {author} {\bibinfo {author} {\bibfnamefont {C.}~\bibnamefont
  {Roberts}}, \bibinfo {author} {\bibfnamefont {E.~M.}\ \bibnamefont
  {Stewart}},\ and\ \bibinfo {author} {\bibfnamefont {F.}~\bibnamefont
  {Milano}},\ }in\ \href {https://doi.org/10.1109/pscc.2016.7540898} {\emph
  {\bibinfo {booktitle} {2016 Power Systems Computation Conference (PSCC)}}}\
  (\bibinfo  {publisher} {IEEE},\ \bibinfo {year} {2016})\ p.\ \bibinfo {pages}
  {1–7}\BibitemShut {NoStop}%
\bibitem [{\citenamefont {Tyloo}\ and\ \citenamefont
  {Jacquod}(2021)}]{tyloo2021primary}%
  \BibitemOpen
  \bibfield  {author} {\bibinfo {author} {\bibfnamefont {M.}~\bibnamefont
  {Tyloo}}\ and\ \bibinfo {author} {\bibfnamefont {P.}~\bibnamefont
  {Jacquod}},\ }\href {https://doi.org/10.1109/lcsys.2020.3006966} {\bibfield
  {journal} {\bibinfo  {journal} {IEEE Control Syst. Lett.}\ }\textbf {\bibinfo
  {volume} {5}},\ \bibinfo {pages} {929–934} (\bibinfo {year}
  {2021})}\BibitemShut {NoStop}%
\bibitem [{\citenamefont {Gorbunov}\ \emph {et~al.}(2022)\citenamefont
  {Gorbunov}, \citenamefont {Peng}, \citenamefont {Bialek},\ and\ \citenamefont
  {Vorobev}}]{gorbunov2022inertia}%
  \BibitemOpen
  \bibfield  {author} {\bibinfo {author} {\bibfnamefont {A.}~\bibnamefont
  {Gorbunov}}, \bibinfo {author} {\bibfnamefont {J.~C.-H.}\ \bibnamefont
  {Peng}}, \bibinfo {author} {\bibfnamefont {J.~W.}\ \bibnamefont {Bialek}},\
  and\ \bibinfo {author} {\bibfnamefont {P.}~\bibnamefont {Vorobev}},\ }\href
  {https://doi.org/10.1109/tpwrs.2022.3147691} {\bibfield  {journal} {\bibinfo
  {journal} {IEEE Transactions on Power Systems}\ }\textbf {\bibinfo {volume}
  {37}},\ \bibinfo {pages} {2459–2462} (\bibinfo {year} {2022})}\BibitemShut
  {NoStop}%
\bibitem [{\citenamefont {Ragwitz}\ and\ \citenamefont
  {Kantz}(2001)}]{Ragwitz2001}%
  \BibitemOpen
  \bibfield  {author} {\bibinfo {author} {\bibfnamefont {M.}~\bibnamefont
  {Ragwitz}}\ and\ \bibinfo {author} {\bibfnamefont {H.}~\bibnamefont
  {Kantz}},\ }\href {https://doi.org/10.1103/physrevlett.87.254501} {\bibfield
  {journal} {\bibinfo  {journal} {Physical Review Letters}\ }\textbf {\bibinfo
  {volume} {87}},\ \bibinfo {pages} {254501} (\bibinfo {year}
  {2001})}\BibitemShut {NoStop}%
\bibitem [{\citenamefont {Friedrich}\ \emph {et~al.}(2002)\citenamefont
  {Friedrich}, \citenamefont {Renner}, \citenamefont {Siefert},\ and\
  \citenamefont {Peinke}}]{Friedrich2002}%
  \BibitemOpen
  \bibfield  {author} {\bibinfo {author} {\bibfnamefont {R.}~\bibnamefont
  {Friedrich}}, \bibinfo {author} {\bibfnamefont {C.}~\bibnamefont {Renner}},
  \bibinfo {author} {\bibfnamefont {M.}~\bibnamefont {Siefert}},\ and\ \bibinfo
  {author} {\bibfnamefont {J.}~\bibnamefont {Peinke}},\ }\href
  {https://doi.org/10.1103/physrevlett.89.149401} {\bibfield  {journal}
  {\bibinfo  {journal} {Physical Review Letters}\ }\textbf {\bibinfo {volume}
  {89}},\ \bibinfo {pages} {149401} (\bibinfo {year} {2002})}\BibitemShut
  {NoStop}%
\bibitem [{\citenamefont {Sura}\ and\ \citenamefont
  {Barsugli}(2002)}]{Sura2002}%
  \BibitemOpen
  \bibfield  {author} {\bibinfo {author} {\bibfnamefont {P.}~\bibnamefont
  {Sura}}\ and\ \bibinfo {author} {\bibfnamefont {J.}~\bibnamefont
  {Barsugli}},\ }\href {https://doi.org/10.1016/s0375-9601(02)01474-3}
  {\bibfield  {journal} {\bibinfo  {journal} {Physics Letters A}\ }\textbf
  {\bibinfo {volume} {305}},\ \bibinfo {pages} {304–311} (\bibinfo {year}
  {2002})}\BibitemShut {NoStop}%
\bibitem [{\citenamefont {Gottschall}\ and\ \citenamefont
  {Peinke}(2008)}]{Gottschall2008}%
  \BibitemOpen
  \bibfield  {author} {\bibinfo {author} {\bibfnamefont {J.}~\bibnamefont
  {Gottschall}}\ and\ \bibinfo {author} {\bibfnamefont {J.}~\bibnamefont
  {Peinke}},\ }\href {https://doi.org/10.1088/1367-2630/10/8/083034} {\bibfield
   {journal} {\bibinfo  {journal} {New Journal of Physics}\ }\textbf {\bibinfo
  {volume} {10}},\ \bibinfo {pages} {083034} (\bibinfo {year}
  {2008})}\BibitemShut {NoStop}%
\bibitem [{\citenamefont {Anteneodo}\ and\ \citenamefont
  {Riera}(2009)}]{Anteneodo2009}%
  \BibitemOpen
  \bibfield  {author} {\bibinfo {author} {\bibfnamefont {C.}~\bibnamefont
  {Anteneodo}}\ and\ \bibinfo {author} {\bibfnamefont {R.}~\bibnamefont
  {Riera}},\ }\href {https://doi.org/10.1103/physreve.80.031103} {\bibfield
  {journal} {\bibinfo  {journal} {Physical Review E}\ }\textbf {\bibinfo
  {volume} {80}},\ \bibinfo {pages} {031103} (\bibinfo {year}
  {2009})}\BibitemShut {NoStop}%
\bibitem [{\citenamefont {Lade}(2009)}]{Lade2009}%
  \BibitemOpen
  \bibfield  {author} {\bibinfo {author} {\bibfnamefont {S.~J.}\ \bibnamefont
  {Lade}},\ }\href {https://doi.org/10.1016/j.physleta.2009.08.029} {\bibfield
  {journal} {\bibinfo  {journal} {Physics Letters A}\ }\textbf {\bibinfo
  {volume} {373}},\ \bibinfo {pages} {3705–3709} (\bibinfo {year}
  {2009})}\BibitemShut {NoStop}%
\bibitem [{\citenamefont {Anteneodo}\ and\ \citenamefont
  {Queirós}(2010)}]{Anteneodo2010}%
  \BibitemOpen
  \bibfield  {author} {\bibinfo {author} {\bibfnamefont {C.}~\bibnamefont
  {Anteneodo}}\ and\ \bibinfo {author} {\bibfnamefont {S.~M.~D.}\ \bibnamefont
  {Queirós}},\ }\href {https://doi.org/10.1103/physreve.82.041122} {\bibfield
  {journal} {\bibinfo  {journal} {Physical Review E}\ }\textbf {\bibinfo
  {volume} {82}},\ \bibinfo {pages} {041122} (\bibinfo {year}
  {2010})}\BibitemShut {NoStop}%
\bibitem [{\citenamefont {Honisch}\ and\ \citenamefont
  {Friedrich}(2011)}]{Honisch2011}%
  \BibitemOpen
  \bibfield  {author} {\bibinfo {author} {\bibfnamefont {C.}~\bibnamefont
  {Honisch}}\ and\ \bibinfo {author} {\bibfnamefont {R.}~\bibnamefont
  {Friedrich}},\ }\href {https://doi.org/10.1103/physreve.83.066701} {\bibfield
   {journal} {\bibinfo  {journal} {Physical Review E}\ }\textbf {\bibinfo
  {volume} {83}},\ \bibinfo {pages} {066701} (\bibinfo {year}
  {2011})}\BibitemShut {NoStop}%
\bibitem [{\citenamefont {Rydin Gorj\~ao}\ \emph {et~al.}(2021)\citenamefont
  {Rydin Gorj\~ao}, \citenamefont {Witthaut}, \citenamefont {Lehnertz},\ and\
  \citenamefont {Lind}}]{RydinGorjao2021}%
  \BibitemOpen
  \bibfield  {author} {\bibinfo {author} {\bibfnamefont {L.}~\bibnamefont
  {Rydin Gorj\~ao}}, \bibinfo {author} {\bibfnamefont {D.}~\bibnamefont
  {Witthaut}}, \bibinfo {author} {\bibfnamefont {K.}~\bibnamefont {Lehnertz}},\
  and\ \bibinfo {author} {\bibfnamefont {P.~G.}\ \bibnamefont {Lind}},\ }\href
  {https://doi.org/10.3390/e23050517} {\bibfield  {journal} {\bibinfo
  {journal} {Entropy}\ }\textbf {\bibinfo {volume} {23}},\ \bibinfo {pages}
  {517} (\bibinfo {year} {2021})}\BibitemShut {NoStop}%
\bibitem [{\citenamefont {B\"{o}ttcher}\ \emph {et~al.}(2006)\citenamefont
  {B\"{o}ttcher}, \citenamefont {Peinke}, \citenamefont {Kleinhans},
  \citenamefont {Friedrich}, \citenamefont {Lind},\ and\ \citenamefont
  {Haase}}]{Boettcher2006}%
  \BibitemOpen
  \bibfield  {author} {\bibinfo {author} {\bibfnamefont {F.}~\bibnamefont
  {B\"{o}ttcher}}, \bibinfo {author} {\bibfnamefont {J.}~\bibnamefont
  {Peinke}}, \bibinfo {author} {\bibfnamefont {D.}~\bibnamefont {Kleinhans}},
  \bibinfo {author} {\bibfnamefont {R.}~\bibnamefont {Friedrich}}, \bibinfo
  {author} {\bibfnamefont {P.~G.}\ \bibnamefont {Lind}},\ and\ \bibinfo
  {author} {\bibfnamefont {M.}~\bibnamefont {Haase}},\ }\href
  {https://doi.org/10.1103/physrevlett.97.090603} {\bibfield  {journal}
  {\bibinfo  {journal} {Physical Review Letters}\ }\textbf {\bibinfo {volume}
  {97}},\ \bibinfo {pages} {090603} (\bibinfo {year} {2006})}\BibitemShut
  {NoStop}%
\bibitem [{\citenamefont {Lehle}(2011)}]{Lehle2011}%
  \BibitemOpen
  \bibfield  {author} {\bibinfo {author} {\bibfnamefont {B.}~\bibnamefont
  {Lehle}},\ }\href {https://doi.org/10.1103/physreve.83.021113} {\bibfield
  {journal} {\bibinfo  {journal} {Physical Review E}\ }\textbf {\bibinfo
  {volume} {83}},\ \bibinfo {pages} {021113} (\bibinfo {year}
  {2011})}\BibitemShut {NoStop}%
\bibitem [{\citenamefont {Lehle}\ and\ \citenamefont
  {Peinke}(2018)}]{Lehle2018}%
  \BibitemOpen
  \bibfield  {author} {\bibinfo {author} {\bibfnamefont {B.}~\bibnamefont
  {Lehle}}\ and\ \bibinfo {author} {\bibfnamefont {J.}~\bibnamefont {Peinke}},\
  }\href {https://doi.org/10.1103/physreve.97.012113} {\bibfield  {journal}
  {\bibinfo  {journal} {Physical Review E}\ }\textbf {\bibinfo {volume} {97}},\
  \bibinfo {pages} {012113} (\bibinfo {year} {2018})}\BibitemShut {NoStop}%
\bibitem [{\citenamefont {Willers}\ and\ \citenamefont
  {Kamps}(2021{\natexlab{a}})}]{Willers2021}%
  \BibitemOpen
  \bibfield  {author} {\bibinfo {author} {\bibfnamefont {C.}~\bibnamefont
  {Willers}}\ and\ \bibinfo {author} {\bibfnamefont {O.}~\bibnamefont
  {Kamps}},\ }\href {https://doi.org/10.1140/epjb/s10051-021-00149-0}
  {\bibfield  {journal} {\bibinfo  {journal} {The European Physical Journal B}\
  }\textbf {\bibinfo {volume} {94}},\ \bibinfo {pages} {149} (\bibinfo {year}
  {2021}{\natexlab{a}})}\BibitemShut {NoStop}%
\bibitem [{\citenamefont {Uhlenbeck}\ and\ \citenamefont
  {Ornstein}(1930)}]{Uhlenbeck1930}%
  \BibitemOpen
  \bibfield  {author} {\bibinfo {author} {\bibfnamefont {G.~E.}\ \bibnamefont
  {Uhlenbeck}}\ and\ \bibinfo {author} {\bibfnamefont {L.~S.}\ \bibnamefont
  {Ornstein}},\ }\href {https://doi.org/10.1103/physrev.36.823} {\bibfield
  {journal} {\bibinfo  {journal} {Physical Review}\ }\textbf {\bibinfo {volume}
  {36}},\ \bibinfo {pages} {823–841} (\bibinfo {year} {1930})}\BibitemShut
  {NoStop}%
\bibitem [{\citenamefont {Fodor}\ \emph {et~al.}(2016)\citenamefont {Fodor},
  \citenamefont {Nardini}, \citenamefont {Cates}, \citenamefont {Tailleur},
  \citenamefont {Visco},\ and\ \citenamefont {van Wijland}}]{Fodor2016}%
  \BibitemOpen
  \bibfield  {author} {\bibinfo {author} {\bibfnamefont {E.}~\bibnamefont
  {Fodor}}, \bibinfo {author} {\bibfnamefont {C.}~\bibnamefont {Nardini}},
  \bibinfo {author} {\bibfnamefont {M.~E.}\ \bibnamefont {Cates}}, \bibinfo
  {author} {\bibfnamefont {J.}~\bibnamefont {Tailleur}}, \bibinfo {author}
  {\bibfnamefont {P.}~\bibnamefont {Visco}},\ and\ \bibinfo {author}
  {\bibfnamefont {F.}~\bibnamefont {van Wijland}},\ }\href
  {https://doi.org/10.1103/physrevlett.117.038103} {\bibfield  {journal}
  {\bibinfo  {journal} {Physical Review Letters}\ }\textbf {\bibinfo {volume}
  {117}},\ \bibinfo {pages} {038103} (\bibinfo {year} {2016})}\BibitemShut
  {NoStop}%
\bibitem [{\citenamefont {Martin}\ \emph {et~al.}(2021)\citenamefont {Martin},
  \citenamefont {O'Byrne}, \citenamefont {Cates}, \citenamefont {Fodor},
  \citenamefont {Nardini}, \citenamefont {Tailleur},\ and\ \citenamefont {van
  Wijland}}]{Martin2021}%
  \BibitemOpen
  \bibfield  {author} {\bibinfo {author} {\bibfnamefont {D.}~\bibnamefont
  {Martin}}, \bibinfo {author} {\bibfnamefont {J.}~\bibnamefont {O'Byrne}},
  \bibinfo {author} {\bibfnamefont {M.~E.}\ \bibnamefont {Cates}}, \bibinfo
  {author} {\bibfnamefont {E.}~\bibnamefont {Fodor}}, \bibinfo {author}
  {\bibfnamefont {C.}~\bibnamefont {Nardini}}, \bibinfo {author} {\bibfnamefont
  {J.}~\bibnamefont {Tailleur}},\ and\ \bibinfo {author} {\bibfnamefont
  {F.}~\bibnamefont {van Wijland}},\ }\href
  {https://doi.org/10.1103/physreve.103.032607} {\bibfield  {journal} {\bibinfo
   {journal} {Physical Review E}\ }\textbf {\bibinfo {volume} {103}},\ \bibinfo
  {pages} {032607} (\bibinfo {year} {2021})}\BibitemShut {NoStop}%
\bibitem [{Note1()}]{Note1}%
  \BibitemOpen
  \bibinfo {note} {See Supplemental Material (yet to be published) for
  derivations, implementation details, and details on the data, including
  references \cite
  {Carlstein1986,Gillespie1996,Willers2021zenodo}.}\BibitemShut {Stop}%
\bibitem [{\citenamefont {Rydin Gorj\~ao}\ \emph {et~al.}(2025)\citenamefont
  {Rydin Gorj\~ao}, \citenamefont {Maritz}, \citenamefont {Sch\"{a}fer},\ and\
  \citenamefont {Witthaut}}]{RydinGorjao2025}%
  \BibitemOpen
  \bibfield  {author} {\bibinfo {author} {\bibfnamefont {L.}~\bibnamefont
  {Rydin Gorj\~ao}}, \bibinfo {author} {\bibfnamefont {J.~M.}\ \bibnamefont
  {Maritz}}, \bibinfo {author} {\bibfnamefont {B.}~\bibnamefont
  {Sch\"{a}fer}},\ and\ \bibinfo {author} {\bibfnamefont {D.}~\bibnamefont
  {Witthaut}},\ }\href {https://doi.org/10.1051/epn/2025510} {\bibfield
  {journal} {\bibinfo  {journal} {Europhysics News}\ }\textbf {\bibinfo
  {volume} {56}},\ \bibinfo {pages} {17–20} (\bibinfo {year}
  {2025})}\BibitemShut {NoStop}%
\bibitem [{\citenamefont {Wen}\ \emph {et~al.}(2025)\citenamefont {Wen},
  \citenamefont {Anvari}, \citenamefont {Rydin Gorj\~ao}, \citenamefont
  {Yalcin}, \citenamefont {Hagenmeyer},\ and\ \citenamefont
  {Sch\"{a}fer}}]{wen2025nonstandard}%
  \BibitemOpen
  \bibfield  {author} {\bibinfo {author} {\bibfnamefont {X.}~\bibnamefont
  {Wen}}, \bibinfo {author} {\bibfnamefont {M.}~\bibnamefont {Anvari}},
  \bibinfo {author} {\bibfnamefont {L.}~\bibnamefont {Rydin Gorj\~ao}},
  \bibinfo {author} {\bibfnamefont {G.~C.}\ \bibnamefont {Yalcin}}, \bibinfo
  {author} {\bibfnamefont {V.}~\bibnamefont {Hagenmeyer}},\ and\ \bibinfo
  {author} {\bibfnamefont {B.}~\bibnamefont {Sch\"{a}fer}},\ }\href
  {https://doi.org/10.1038/s41598-025-25334-4} {\bibfield  {journal} {\bibinfo
  {journal} {Scientific Reports}\ }\textbf {\bibinfo {volume} {15}},\ \bibinfo
  {pages} {38470} (\bibinfo {year} {2025})}\BibitemShut {NoStop}%
\bibitem [{\citenamefont {Honisch}\ \emph {et~al.}(2012)\citenamefont
  {Honisch}, \citenamefont {Friedrich}, \citenamefont {H\"{o}rner},\ and\
  \citenamefont {Denz}}]{Honisch2012}%
  \BibitemOpen
  \bibfield  {author} {\bibinfo {author} {\bibfnamefont {C.}~\bibnamefont
  {Honisch}}, \bibinfo {author} {\bibfnamefont {R.}~\bibnamefont {Friedrich}},
  \bibinfo {author} {\bibfnamefont {F.}~\bibnamefont {H\"{o}rner}},\ and\
  \bibinfo {author} {\bibfnamefont {C.}~\bibnamefont {Denz}},\ }\href
  {https://doi.org/10.1103/physreve.86.026702} {\bibfield  {journal} {\bibinfo
  {journal} {Physical Review E}\ }\textbf {\bibinfo {volume} {86}},\ \bibinfo
  {pages} {026702} (\bibinfo {year} {2012})}\BibitemShut {NoStop}%
\bibitem [{\citenamefont {L\"{u}ck}\ \emph {et~al.}(2006)\citenamefont
  {L\"{u}ck}, \citenamefont {Renner}, \citenamefont {Peinke},\ and\
  \citenamefont {Friedrich}}]{Lueck2006}%
  \BibitemOpen
  \bibfield  {author} {\bibinfo {author} {\bibfnamefont {S.}~\bibnamefont
  {L\"{u}ck}}, \bibinfo {author} {\bibfnamefont {C.}~\bibnamefont {Renner}},
  \bibinfo {author} {\bibfnamefont {J.}~\bibnamefont {Peinke}},\ and\ \bibinfo
  {author} {\bibfnamefont {R.}~\bibnamefont {Friedrich}},\ }\href
  {https://doi.org/10.1016/j.physleta.2006.06.053} {\bibfield  {journal}
  {\bibinfo  {journal} {Physics Letters A}\ }\textbf {\bibinfo {volume}
  {359}},\ \bibinfo {pages} {335–338} (\bibinfo {year} {2006})}\BibitemShut
  {NoStop}%
\bibitem [{\citenamefont {Stresing}\ \emph {et~al.}(2011)\citenamefont
  {Stresing}, \citenamefont {Kleinhans}, \citenamefont {Friedrich},\ and\
  \citenamefont {Peinke}}]{Stresing2011}%
  \BibitemOpen
  \bibfield  {author} {\bibinfo {author} {\bibfnamefont {R.}~\bibnamefont
  {Stresing}}, \bibinfo {author} {\bibfnamefont {D.}~\bibnamefont {Kleinhans}},
  \bibinfo {author} {\bibfnamefont {R.}~\bibnamefont {Friedrich}},\ and\
  \bibinfo {author} {\bibfnamefont {J.}~\bibnamefont {Peinke}},\ }\href
  {https://doi.org/10.1103/physreve.83.046319} {\bibfield  {journal} {\bibinfo
  {journal} {Physical Review E}\ }\textbf {\bibinfo {volume} {83}},\ \bibinfo
  {pages} {046319} (\bibinfo {year} {2011})}\BibitemShut {NoStop}%
\bibitem [{Note2()}]{Note2}%
  \BibitemOpen
  \bibinfo {note} {Power-grid frequency data for 2023 downloaded from NESO
  (National Energy System Operator, GB), \protect \href
  {https://www.neso.energy/data-portal/system-frequency-data}{https://www.neso.energy/data-portal/system-frequency-data},
  accessed 9 Sept 2026}\BibitemShut {NoStop}%
\bibitem [{\citenamefont
  {Onsager}(1931{\natexlab{a}})}]{onsager1931reciprocal1}%
  \BibitemOpen
  \bibfield  {author} {\bibinfo {author} {\bibfnamefont {L.}~\bibnamefont
  {Onsager}},\ }\href {https://doi.org/10.1103/physrev.37.405} {\bibfield
  {journal} {\bibinfo  {journal} {Physical Review}\ }\textbf {\bibinfo {volume}
  {37}},\ \bibinfo {pages} {405–426} (\bibinfo {year}
  {1931}{\natexlab{a}})}\BibitemShut {NoStop}%
\bibitem [{\citenamefont
  {Onsager}(1931{\natexlab{b}})}]{onsager1931reciprocal2}%
  \BibitemOpen
  \bibfield  {author} {\bibinfo {author} {\bibfnamefont {L.}~\bibnamefont
  {Onsager}},\ }\href {https://doi.org/10.1103/physrev.38.2265} {\bibfield
  {journal} {\bibinfo  {journal} {Physical Review}\ }\textbf {\bibinfo {volume}
  {38}},\ \bibinfo {pages} {2265–2279} (\bibinfo {year}
  {1931}{\natexlab{b}})}\BibitemShut {NoStop}%
\bibitem [{\citenamefont {Reimann}\ and\ \citenamefont
  {Eidecker-Dunkel}(2024)}]{reimann2024onsager}%
  \BibitemOpen
  \bibfield  {author} {\bibinfo {author} {\bibfnamefont {P.}~\bibnamefont
  {Reimann}}\ and\ \bibinfo {author} {\bibfnamefont {C.}~\bibnamefont
  {Eidecker-Dunkel}},\ }\href {https://doi.org/10.1103/physrevb.110.014306}
  {\bibfield  {journal} {\bibinfo  {journal} {Physical Review B}\ }\textbf
  {\bibinfo {volume} {110}},\ \bibinfo {pages} {014306} (\bibinfo {year}
  {2024})}\BibitemShut {NoStop}%
\bibitem [{\citenamefont {Lehle}\ and\ \citenamefont
  {Peinke}(2015)}]{Lehle2015}%
  \BibitemOpen
  \bibfield  {author} {\bibinfo {author} {\bibfnamefont {B.}~\bibnamefont
  {Lehle}}\ and\ \bibinfo {author} {\bibfnamefont {J.}~\bibnamefont {Peinke}},\
  }\href {https://doi.org/10.1103/physreve.91.062113} {\bibfield  {journal}
  {\bibinfo  {journal} {Physical Review E}\ }\textbf {\bibinfo {volume} {91}},\
  \bibinfo {pages} {062113} (\bibinfo {year} {2015})}\BibitemShut {NoStop}%
\bibitem [{\citenamefont {Dembo}\ and\ \citenamefont
  {Zeitouni}(1986)}]{Dembo1986}%
  \BibitemOpen
  \bibfield  {author} {\bibinfo {author} {\bibfnamefont {A.}~\bibnamefont
  {Dembo}}\ and\ \bibinfo {author} {\bibfnamefont {O.}~\bibnamefont
  {Zeitouni}},\ }\href {https://doi.org/10.1016/0304-4149(86)90018-9}
  {\bibfield  {journal} {\bibinfo  {journal} {Stochastic Processes and their
  Applications}\ }\textbf {\bibinfo {volume} {23}},\ \bibinfo {pages}
  {91–113} (\bibinfo {year} {1986})}\BibitemShut {NoStop}%
\bibitem [{\citenamefont {Campillo}\ and\ \citenamefont
  {Le~Gland}(1989)}]{Campillo1989}%
  \BibitemOpen
  \bibfield  {author} {\bibinfo {author} {\bibfnamefont {F.}~\bibnamefont
  {Campillo}}\ and\ \bibinfo {author} {\bibfnamefont {F.}~\bibnamefont
  {Le~Gland}},\ }\href {https://doi.org/10.1016/0304-4149(89)90041-0}
  {\bibfield  {journal} {\bibinfo  {journal} {Stochastic Processes and their
  Applications}\ }\textbf {\bibinfo {volume} {33}},\ \bibinfo {pages}
  {245–274} (\bibinfo {year} {1989})}\BibitemShut {NoStop}%
\bibitem [{\citenamefont {Golightly}\ and\ \citenamefont
  {Wilkinson}(2008)}]{Golightly2008}%
  \BibitemOpen
  \bibfield  {author} {\bibinfo {author} {\bibfnamefont {A.}~\bibnamefont
  {Golightly}}\ and\ \bibinfo {author} {\bibfnamefont {D.}~\bibnamefont
  {Wilkinson}},\ }\href {https://doi.org/10.1016/j.csda.2007.05.019} {\bibfield
   {journal} {\bibinfo  {journal} {Comput. Stat. Data Anal.}\ }\textbf
  {\bibinfo {volume} {52}},\ \bibinfo {pages} {1674–1693} (\bibinfo {year}
  {2008})}\BibitemShut {NoStop}%
\bibitem [{\citenamefont {Rydin Gorj\~ao}\ and\ \citenamefont
  {Meirinhos}(2019)}]{Gorjao2019}%
  \BibitemOpen
  \bibfield  {author} {\bibinfo {author} {\bibfnamefont {L.}~\bibnamefont
  {Rydin Gorj\~ao}}\ and\ \bibinfo {author} {\bibfnamefont {F.}~\bibnamefont
  {Meirinhos}},\ }\href {https://doi.org/10.21105/joss.01693} {\bibfield
  {journal} {\bibinfo  {journal} {Journal of Open Source Software}\ }\textbf
  {\bibinfo {volume} {4}},\ \bibinfo {pages} {1693} (\bibinfo {year}
  {2019})}\BibitemShut {NoStop}%
\bibitem [{\citenamefont {Basmann}(1957)}]{Basmann1957}%
  \BibitemOpen
  \bibfield  {author} {\bibinfo {author} {\bibfnamefont {R.~L.}\ \bibnamefont
  {Basmann}},\ }\href {https://doi.org/10.2307/1907743} {\bibfield  {journal}
  {\bibinfo  {journal} {Econometrica}\ }\textbf {\bibinfo {volume} {25}},\
  \bibinfo {pages} {77} (\bibinfo {year} {1957})}\BibitemShut {NoStop}%
\bibitem [{\citenamefont {Sargan}(1958)}]{Sargan1958}%
  \BibitemOpen
  \bibfield  {author} {\bibinfo {author} {\bibfnamefont {J.~D.}\ \bibnamefont
  {Sargan}},\ }\href {https://doi.org/10.2307/1907619} {\bibfield  {journal}
  {\bibinfo  {journal} {Econometrica}\ }\textbf {\bibinfo {volume} {26}},\
  \bibinfo {pages} {393} (\bibinfo {year} {1958})}\BibitemShut {NoStop}%
\bibitem [{\citenamefont {Angrist}\ and\ \citenamefont
  {Krueger}(2001)}]{Angrist2001}%
  \BibitemOpen
  \bibfield  {author} {\bibinfo {author} {\bibfnamefont {J.~D.}\ \bibnamefont
  {Angrist}}\ and\ \bibinfo {author} {\bibfnamefont {A.~B.}\ \bibnamefont
  {Krueger}},\ }\href {https://doi.org/10.1257/jep.15.4.69} {\bibfield
  {journal} {\bibinfo  {journal} {Journal of Economic Perspectives}\ }\textbf
  {\bibinfo {volume} {15}},\ \bibinfo {pages} {69–85} (\bibinfo {year}
  {2001})}\BibitemShut {NoStop}%
\bibitem [{\citenamefont {Hayashi}(2000)}]{Hayashi2000}%
  \BibitemOpen
  \bibfield  {author} {\bibinfo {author} {\bibfnamefont {F.}~\bibnamefont
  {Hayashi}},\ }\href
  {https://press.princeton.edu/books/hardcover/9780691010182/econometrics?srsltid=AU7gw4WUbF3jREJZixmujnvihq5Glu54FMnGTqNkIRk9gymI5PcnPAlO}
  {\emph {\bibinfo {title} {Econometrics}}}\ (\bibinfo  {publisher} {Princeton
  University Press},\ \bibinfo {address} {Princeton, NJ},\ \bibinfo {year}
  {2000})\BibitemShut {NoStop}%
\bibitem [{\citenamefont {Hansen}(1982)}]{Hansen1982}%
  \BibitemOpen
  \bibfield  {author} {\bibinfo {author} {\bibfnamefont {L.~P.}\ \bibnamefont
  {Hansen}},\ }\href {https://doi.org/10.2307/1912775} {\bibfield  {journal}
  {\bibinfo  {journal} {Econometrica}\ }\textbf {\bibinfo {volume} {50}},\
  \bibinfo {pages} {1029} (\bibinfo {year} {1982})}\BibitemShut {NoStop}%
\bibitem [{\citenamefont {Hansen}\ and\ \citenamefont
  {Singleton}(1982)}]{HansenSingleton1982}%
  \BibitemOpen
  \bibfield  {author} {\bibinfo {author} {\bibfnamefont {L.~P.}\ \bibnamefont
  {Hansen}}\ and\ \bibinfo {author} {\bibfnamefont {K.~J.}\ \bibnamefont
  {Singleton}},\ }\href {https://doi.org/10.2307/1911873} {\bibfield  {journal}
  {\bibinfo  {journal} {Econometrica}\ }\textbf {\bibinfo {volume} {50}},\
  \bibinfo {pages} {1269} (\bibinfo {year} {1982})}\BibitemShut {NoStop}%
\bibitem [{\citenamefont {Hall}(2010)}]{Hall2010}%
  \BibitemOpen
  \bibfield  {author} {\bibinfo {author} {\bibfnamefont {A.~R.}\ \bibnamefont
  {Hall}},\ }\href {https://doi.org/10.1002/9780470061602.eqf19004} {\bibinfo
  {title} {Generalized method of moments ({GMM})}} (\bibinfo {year}
  {2010})\BibitemShut {NoStop}%
\bibitem [{\citenamefont {Nikakhtar}\ \emph {et~al.}(2023)\citenamefont
  {Nikakhtar}, \citenamefont {Parkavousi}, \citenamefont {Sahimi},
  \citenamefont {Reza Rahimi~Tabar}, \citenamefont {Feudel},\ and\
  \citenamefont {Lehnertz}}]{Nikakhtar2023}%
  \BibitemOpen
  \bibfield  {author} {\bibinfo {author} {\bibfnamefont {F.}~\bibnamefont
  {Nikakhtar}}, \bibinfo {author} {\bibfnamefont {L.}~\bibnamefont
  {Parkavousi}}, \bibinfo {author} {\bibfnamefont {M.}~\bibnamefont {Sahimi}},
  \bibinfo {author} {\bibfnamefont {M.}~\bibnamefont {Reza Rahimi~Tabar}},
  \bibinfo {author} {\bibfnamefont {U.}~\bibnamefont {Feudel}},\ and\ \bibinfo
  {author} {\bibfnamefont {K.}~\bibnamefont {Lehnertz}},\ }\href
  {https://doi.org/10.1088/1367-2630/acec63} {\bibfield  {journal} {\bibinfo
  {journal} {New Journal of Physics}\ }\textbf {\bibinfo {volume} {25}},\
  \bibinfo {pages} {083025} (\bibinfo {year} {2023})}\BibitemShut {NoStop}%
\bibitem [{\citenamefont {Shinoda}\ \emph {et~al.}(2023)\citenamefont
  {Shinoda}, \citenamefont {Dai}, \citenamefont {Bakhos}, \citenamefont
  {Gonzalez-Torres}, \citenamefont {Benchaib},\ and\ \citenamefont
  {Bacha}}]{Shinoda2023}%
  \BibitemOpen
  \bibfield  {author} {\bibinfo {author} {\bibfnamefont {K.}~\bibnamefont
  {Shinoda}}, \bibinfo {author} {\bibfnamefont {J.}~\bibnamefont {Dai}},
  \bibinfo {author} {\bibfnamefont {G.}~\bibnamefont {Bakhos}}, \bibinfo
  {author} {\bibfnamefont {J.~C.}\ \bibnamefont {Gonzalez-Torres}}, \bibinfo
  {author} {\bibfnamefont {A.}~\bibnamefont {Benchaib}},\ and\ \bibinfo
  {author} {\bibfnamefont {S.}~\bibnamefont {Bacha}},\ }\href
  {https://doi.org/10.1049/gtd2.12955} {\bibfield  {journal} {\bibinfo
  {journal} {IET Generation, Transmission \& Distribution}\ }\textbf {\bibinfo
  {volume} {17}},\ \bibinfo {pages} {4024–4037} (\bibinfo {year}
  {2023})}\BibitemShut {NoStop}%
\bibitem [{\citenamefont {Shouran}\ \emph {et~al.}(2021)\citenamefont
  {Shouran}, \citenamefont {Anayi}, \citenamefont {Packianather},\ and\
  \citenamefont {Habil}}]{Shouran2021}%
  \BibitemOpen
  \bibfield  {author} {\bibinfo {author} {\bibfnamefont {M.}~\bibnamefont
  {Shouran}}, \bibinfo {author} {\bibfnamefont {F.}~\bibnamefont {Anayi}},
  \bibinfo {author} {\bibfnamefont {M.}~\bibnamefont {Packianather}},\ and\
  \bibinfo {author} {\bibfnamefont {M.}~\bibnamefont {Habil}},\ }\href
  {https://doi.org/10.3390/designs5030050} {\bibfield  {journal} {\bibinfo
  {journal} {Designs}\ }\textbf {\bibinfo {volume} {5}},\ \bibinfo {pages} {50}
  (\bibinfo {year} {2021})}\BibitemShut {NoStop}%
\bibitem [{\citenamefont {Vorobev}\ \emph {et~al.}(2019)\citenamefont
  {Vorobev}, \citenamefont {Greenwood}, \citenamefont {Bell}, \citenamefont
  {Bialek}, \citenamefont {Taylor},\ and\ \citenamefont
  {Turitsyn}}]{vorobev2019deadbands}%
  \BibitemOpen
  \bibfield  {author} {\bibinfo {author} {\bibfnamefont {P.}~\bibnamefont
  {Vorobev}}, \bibinfo {author} {\bibfnamefont {D.~M.}\ \bibnamefont
  {Greenwood}}, \bibinfo {author} {\bibfnamefont {J.~H.}\ \bibnamefont {Bell}},
  \bibinfo {author} {\bibfnamefont {J.~W.}\ \bibnamefont {Bialek}}, \bibinfo
  {author} {\bibfnamefont {P.~C.}\ \bibnamefont {Taylor}},\ and\ \bibinfo
  {author} {\bibfnamefont {K.}~\bibnamefont {Turitsyn}},\ }\href
  {https://doi.org/10.1109/TPWRS.2019.2895547} {\bibfield  {journal} {\bibinfo
  {journal} {IEEE Transactions on Power Systems}\ }\textbf {\bibinfo {volume}
  {34}},\ \bibinfo {pages} {3098} (\bibinfo {year} {2019})}\BibitemShut
  {NoStop}%
\bibitem [{\citenamefont {del Giudice}\ \emph {et~al.}(2021)\citenamefont {del
  Giudice}, \citenamefont {Brambilla}, \citenamefont {Grillo},\ and\
  \citenamefont {Bizzarri}}]{delgiudice2021effects}%
  \BibitemOpen
  \bibfield  {author} {\bibinfo {author} {\bibfnamefont {D.}~\bibnamefont {del
  Giudice}}, \bibinfo {author} {\bibfnamefont {A.}~\bibnamefont {Brambilla}},
  \bibinfo {author} {\bibfnamefont {S.}~\bibnamefont {Grillo}},\ and\ \bibinfo
  {author} {\bibfnamefont {F.}~\bibnamefont {Bizzarri}},\ }\href
  {https://doi.org/10.1016/j.ijepes.2021.106842} {\bibfield  {journal}
  {\bibinfo  {journal} {Int. J. Electr. Power Energy Syst.}\ }\textbf {\bibinfo
  {volume} {129}},\ \bibinfo {pages} {106842} (\bibinfo {year}
  {2021})}\BibitemShut {NoStop}%
\bibitem [{\citenamefont {Anvari}\ \emph {et~al.}(2016)\citenamefont {Anvari},
  \citenamefont {Lohmann}, \citenamefont {W\"{a}chter}, \citenamefont {Milan},
  \citenamefont {Lorenz}, \citenamefont {Heinemann}, \citenamefont {Tabar},\
  and\ \citenamefont {Peinke}}]{Anvari2016Short}%
  \BibitemOpen
  \bibfield  {author} {\bibinfo {author} {\bibfnamefont {M.}~\bibnamefont
  {Anvari}}, \bibinfo {author} {\bibfnamefont {G.}~\bibnamefont {Lohmann}},
  \bibinfo {author} {\bibfnamefont {M.}~\bibnamefont {W\"{a}chter}}, \bibinfo
  {author} {\bibfnamefont {P.}~\bibnamefont {Milan}}, \bibinfo {author}
  {\bibfnamefont {E.}~\bibnamefont {Lorenz}}, \bibinfo {author} {\bibfnamefont
  {D.}~\bibnamefont {Heinemann}}, \bibinfo {author} {\bibfnamefont {M.~R.~R.}\
  \bibnamefont {Tabar}},\ and\ \bibinfo {author} {\bibfnamefont
  {J.}~\bibnamefont {Peinke}},\ }\href
  {https://doi.org/10.1088/1367-2630/18/6/063027} {\bibfield  {journal}
  {\bibinfo  {journal} {New Journal of Physics}\ }\textbf {\bibinfo {volume}
  {18}},\ \bibinfo {pages} {063027} (\bibinfo {year} {2016})}\BibitemShut
  {NoStop}%
\bibitem [{\citenamefont
  {Villalobos~Concha}(2025)}]{VillalobosConcha2025Zenodo}%
  \BibitemOpen
  \bibfield  {author} {\bibinfo {author} {\bibfnamefont {C.}~\bibnamefont
  {Villalobos~Concha}},\ }\href {https://doi.org/10.5281/ZENODO.15881570}
  {\bibinfo {title} {Active bacterial bath in droplets}} (\bibinfo {year}
  {2025})\BibitemShut {NoStop}%
\bibitem [{\citenamefont {Carlstein}(1986)}]{Carlstein1986}%
  \BibitemOpen
  \bibfield  {author} {\bibinfo {author} {\bibfnamefont {E.}~\bibnamefont
  {Carlstein}},\ }\href {https://doi.org/10.1214/aos/1176350057} {\bibfield
  {journal} {\bibinfo  {journal} {Ann. Statist.}\ }\textbf {\bibinfo {volume}
  {14}},\ \bibinfo {pages} {1171} (\bibinfo {year} {1986})}\BibitemShut
  {NoStop}%
\bibitem [{\citenamefont {Gillespie}(1996)}]{Gillespie1996}%
  \BibitemOpen
  \bibfield  {author} {\bibinfo {author} {\bibfnamefont {D.~T.}\ \bibnamefont
  {Gillespie}},\ }\href {https://doi.org/10.1103/physreve.54.2084} {\bibfield
  {journal} {\bibinfo  {journal} {Physical Review E}\ }\textbf {\bibinfo
  {volume} {54}},\ \bibinfo {pages} {2084–2091} (\bibinfo {year}
  {1996})}\BibitemShut {NoStop}%
\bibitem [{\citenamefont {Willers}\ and\ \citenamefont
  {Kamps}(2021{\natexlab{b}})}]{Willers2021zenodo}%
  \BibitemOpen
  \bibfield  {author} {\bibinfo {author} {\bibfnamefont {C.}~\bibnamefont
  {Willers}}\ and\ \bibinfo {author} {\bibfnamefont {O.}~\bibnamefont
  {Kamps}},\ }\href {https://doi.org/10.5281/ZENODO.4986257} {\bibinfo {title}
  {Exemplary python implementation of the estimation procedures proposed in
  "non-parametric estimation of a langevin model driven by correlated noise"}}
  (\bibinfo {year} {2021}{\natexlab{b}})\BibitemShut {NoStop}%
\end{thebibliography}%
\end{document}